\documentclass{emulateapj}
\usepackage{apjfonts}

\usepackage{ifpdf}
\ifpdf
\else
\fi

\begin{document}
\slugcomment{(Accepted for publication in \textit{The Astronomical Journal} - Revised: September 8, 2011)}

\shorttitle{THE TIDAL STREAM OF MESSIER 63}
\shortauthors{T. S. CHONIS ET AL.}

\title{A Petal of the Sunflower: \\ Photometry of the Stellar Tidal Stream in the Halo of Messier 63 (NGC 5055)\altaffilmark{*}}

\author{Taylor S. Chonis\altaffilmark{1}, David Mart\'{i}nez-Delgado\altaffilmark{2}, R. Jay Gabany\altaffilmark{3},\\ Steven R. Majewski\altaffilmark{4}, Gary J. Hill\altaffilmark{5}, Ray Gralak\altaffilmark{6}, and Ignacio Trujillo\altaffilmark{7,}\altaffilmark{8,}\altaffilmark{9}}

\altaffiltext{*} {This paper includes data taken at The McDonald Observatory of The University of Texas at Austin.}
\altaffiltext{1} {Dept. of Astronomy, University of Texas at Austin, 1 University Station, C1400, Austin, TX 78712, USA: tschonis@astro.as.utexas.edu}
\altaffiltext{2} {Max-Planck Institut f\"ur Astronomie, K\"{o}nigstuhl 17, D-69117 Heidelberg, Germany}
\altaffiltext{3} {BlackBird Observatory, Mayhill, New Mexico, USA}
\altaffiltext{4} {Dept. of Astronomy, University of Virginia, 530 McCormick Rd., Charlottesville, VA 22904, USA}
\altaffiltext{5} {McDonald Observatory, University of Texas at Austin, 1 University Station, C1402, Austin, TX 78712, USA}
\altaffiltext{6} {Sirius Imaging Observatory, Mayhill, New Mexico, USA}
\altaffiltext{7} {Instituto de Astrof\'isica de Canarias, C/ V\'{i}a L\'{a}ctea, s/n, E38205 - La Laguna (Tenerife), Spain}
\altaffiltext{8} {Departamento de Astrof\'isica, Universidad de La Laguna, La Laguna (Tenerife), Spain}
\altaffiltext{9} {Ram\'{o}n y Cajal Fellow}

\begin{abstract}
We present deep surface photometry of a very faint, giant arc-loop feature in the halo of the nearby spiral galaxy NGC 5055 (M63) that is consistent with being a part of a stellar stream resulting from the disruption of a dwarf satellite galaxy. This faint feature was first detected in early photographic studies by \citet{vanderKruit1979}; more recently by \citet{martinezDelgado2010} and as presented in this work, the loop has been realized to be the result of a recent minor merger through evidence obtained by wide-field, deep images taken with a telescope of only 0.16 m aperture. The stellar stream is clearly confirmed in additional deep images taken with the 0.5 m telescope of the BlackBird Remote Observatory and the 0.8 m telescope of the McDonald Observatory. This low surface brightness ($\mu_{R} \approx$ 26 mag arcsec$^{-2}$), arc-like structure around the disk of the galaxy extends 14.0$\arcmin$ ($\sim$29 kpc projected) from its center, with a projected width of 1.6$\arcmin$ ($\sim$3.3 kpc). The stream's morphology is consistent with that of the visible part of a giant, ``great-circle" type stellar stream originating from the recent accretion of a $\sim$10$^{8}$ M$_{\odot}$ dwarf satellite in the last few Gyr. The progenitor satellite's current position and final fate are not conclusive from our data. The color of the stream's stars is consistent with dwarfs in the Local Group and is similar to the outer faint regions of M63's disk and stellar halo. Through our photometric study, we detect other low surface brightness ``plumes"; some of these may be extended spiral features related to the galaxy's complex spiral structure and others may be tidal debris associated with the disruption of the galaxy's outer stellar disk as a result of the accretion event. We are able to differentiate between features related to the tidal stream and faint, blue, extended features in the outskirts of the galaxy's disk previously detected by the GALEX satellite. With its highly warped HI gaseous disk ($\sim$20$^{\circ}$), M63 represents one of several examples of an isolated spiral galaxy with a warped disk showing recently discovered strong evidence of an ongoing minor merger. 
\\ 
\end{abstract}
\keywords{galaxies: dwarf --- galaxies: evolution --- galaxies: halos --- galaxies: individual (NGC 5505) --- galaxies: interactions --- galaxies: photometry}

\section{INTRODUCTION}\label{sec:Introduction}
In the context of a cold dark matter ($\Lambda$CDM) universe, dark matter halo mergers (and the subsequent merging of their baryonic components) drive the evolution of galaxies \citep{white1991}. While state of the art cosmological simulations now show this (e.g., \citealp{springel2005}), it has been advanced for nearly 40 years that galaxy mergers create a hierarchical framework describing galactic evolution, based in part on the characterization of stellar populations in the Milky-Way (MW) stellar halo \citep{searlezinn1978} and on views of extra-galactic interactions \citep{toomre1972}. While major galactic mergers are rare in $\Lambda$CDM at the present epoch \citep{robaina2009}, minor mergers (i.e., those occurring between a massive galaxy and a much less massive satellite) are expected to occur continuously and play an influential role in the ongoing evolution of present day galaxies. Minor mergers where the more massive galaxy is a spiral generally leaves pre-existing stellar disks intact \citep{robertson2006}, though often dynamically altered \citep{velazquezwhite1999}. Even if no obvious interacting satellite can be found, signatures of such an interaction may be present for several Gyr, such as optical or HI disk warps \citep{sancisi1976} or heated disks \citep{tothostriker1992, hernquistquinn1993, velazquezwhite1999}. In a $\Lambda$CDM cosmology, the heating and subsequent survival of disks through many minor mergers over cosmic time remains a debated topic of great interest (e.g., \citealp{purcell2009, moster2010}). Recent simulations (e.g., \citealp{bullock2005, johnston2008, cooper2010}) have shown that another observable signature of such minor mergers should be found in the form of low surface brightness tidal features (e.g., stellar streams) resulting from the disruption of the satellite while still in orbit. These features extend into the stellar halo of the parent spiral and could be viewed as evidence of the evolution of a spiral galaxy stellar halo in-action, an event postulated as long ago as \citet{searlezinn1978}.  

With the advent of wide-field digital sky surveys, such substructure has been found in the stellar halo of the MW in the form of faint over-densities and streams that are resolved into individual stars and reveal the evolutionary history of our galaxy. Arguably, the most spectacular of such structures is the well-studied Sagittarius dwarf galaxy and its tidal stream (e.g., \citealp{majewski2003}), which has been found to be quite complex and rich in substructure \citep{belokurov2006}. This system has been extensively modeled (e.g., \citealp{law2010}, \citealp{penarrubia2010a}). Observational evidence shows that its tidal tails appear to consist of long streams of debris wrapping around the MW in a near polar, rosette-like orbit \citep{martinezDelgado2004}. A wealth of other streams have also been detected surrounding our galaxy, such as the low galactic latitude Monoceros tidal stream \citep{yanny2003}, the Orphan stream \citep{belokurov2007}, and the Anticenter stream \citep{grillmair2006}. Similar structures have also been observed around our nearest massive neighbor, the Andromeda Galaxy (M31; \citealp{ibata2001}, \citealp{mcconnachie2009}), including giant features extending $>$200 kpc towards M33 \citep{ibata2007, mcconnachie2010}. Such features are evidence for the inside-out formation of galaxies and the role of minor mergers in the evolution of spiral galaxy stellar halos in the Local Group. 

If such substructure can be observed around other nearby, relatively isolated galaxies outside the Local Group, this conclusion could be generalized to late-type massive spirals in the present epoch. This prospect has been explored through the semi-analytical models of \citet{johnston2001} and more recently by the $\Lambda$CDM-based models of \citet{bullock2005}, \citet{johnston2008}, and \citet{cooper2010}. In fact, these authors predict that in a statistically significant sample of isolated late-type Local Volume spirals, nearly every galaxy should display such a fossil record of its recent evolution if imaged to sufficient depth (e.g., $\mu_{V} \approx 30$ mag arcsec$^{-2}$). Indeed, several examples of extra-galactic tidal streams resulting from dwarf accretion currently exist as a result of recent research. \citet{malin1997} have performed deep imaging on many galaxies and found loop-like features around M83 and M104. A similar loop structure was discovered by \citet{shang1998} around NGC 5907, the prototypical isolated galaxy displaying a significant optical and HI warp. Deep imaging of NGC 5907 by \citet{martinezDelgado2008} later revealed a stunning complex of streams in addition to the brightest one discovered by \citet{shang1998}. \citet{martinezDelgado2009} discovered a stream resembling models of the MW's Monoceros stellar stream \citep{yanny2003} around NGC 4013, which displays one of the largest known HI warps despite its relative isolation. \citet{wehner2005} discovered a set of streamers indicating a complex merger history around the face-on starburst galaxy NGC 3310 and later performed a detailed study to find that the material did not originate from NGC 3310 itself \citep{wehner2006}. Recently, \citet{mouhcine2010} discovered streams and a thick stellar envelope around NGC 891, a MW analogue, and performed the first ground-based resolved study of stars in a tidal stream system outside of the Local Group. Despite these examples, there are still relatively few known tidal stream systems outside of the Local Group.   

Observing extra-galactic stellar streams due to dwarf accretion analogous to those found in the Local Group can show us that our own galaxy is not unusual, reveal different phases of such interactions, and can open up discussions of different mass and orbit combinations and accretion histories. More fundamentally, it will provide statistical tests of the predictions of galaxy formation and evolutionary models based on the $\Lambda$CDM paradigm. The stunning perspective of extra-galactic tidal streams obtained with modest instruments like those used by \citet{martinezDelgado2008, martinezDelgado2009} for NGC 5907 and NGC 4013, respectively, encourages a more systematic look for these ghostly structures in the nearby universe. With that purpose, a pilot survey was conducted of stellar tidal streams in a select sample of nearby, MW-like spiral systems using modest aperture telescopes (0.1 - 0.5 m) operating under very dark skies \citep{martinezDelgado2010}. For the first time, this pilot survey has allowed a comparison to $\Lambda$CDM based simulations that model the evolution of stellar halos through satellite disruption (see the cited works in the preceding paragraph) with observations. There exists amazing morphological agreement between these simulations and the range of tidal features observed around nearby ``normal" disk galaxies.  

Here, we present surface photometry of similar low surface brightness features around NGC 5055 (hereafter, M63), including a large loop-like structure that is consistent with being part of a stellar tidal stream displaying a ``great-circle" morphology \citep{johnston2008, martinezDelgado2010}. M63 is a large, relatively isolated SA(rs)bc galaxy \citep{deVaucouleurs1991} situated at a distance of 7.2 Mpc \citep{pierce1994}, and belongs to the M51 galaxy group \citep{fouque1992}. Its optical disk displays a fragmented and patchy pattern of spiral arms that extend outwards, resembling a celestial flower (hence its popular name: ``The Sunflower Galaxy"). This characteristic places it in the class of flocculent spirals \citep{elmegreen1987}. Similar to discoveries made for NGC 5907 and NGC 4013, a recent HI study of M63 by \citet{battaglia2006} revealed a \textit{very} pronounced warp ($\sim$20$^{\circ}$) in its gaseous disk that extends out $\sim$40 kpc from its center. Additionally, M63 has recently been found through a GALEX study by \citet{thilker2007} to host a Type 1 extended UV (XUV) disk that is characterized by significant star formation well beyond the anticipated star formation threshold (i.e., the H$\alpha$ edge; see \citealp{martin2001}) at a galactocentric distance of up to $\sim$20 kpc. According to the HYPERLEDA database, M63 has a total $B$ band absolute magnitude of -21.2 AB mag and a radius of 11.7\arcmin\ measured at the $\mu_{B}$ = 25 AB mag arcsec$^{-2}$ isophote. For the reader's reference, we note that the stellar mass and total dark matter halo mass of M63 within 40 kpc are 8$\times$10$^{10}$ M$_{\odot}$ and 2$\times$10$^{11}$ M$_{\odot}$, respectively, as determined by \citet{battaglia2006} by fitting the observed rotation curve with a model that includes a ``maximum disk'' for the stellar component and an isothermal halo.

The faint arc-like structure in the outskirts of M63 was first detected and speculated upon in a photographic study by \citet{vanderKruit1979}. The current work will show that this structure is consistent with being part of the ongoing evolution of M63's stellar halo. Along with the strong HI warp, this feature is evidence for a recent interaction with a low-mass companion and becomes yet another example of the possible link between disk warps and recent mergers. In the following section, we describe the observations, data reduction, and analysis in detail. In $\S$\ref{sec:LSBFeatures}, we discuss the myriad of low surface brightness features detected and measured in our deep images. Finally, through various subsections in $\S$\ref{sec:Discussion}, we outline and discuss our general conclusions. Throughout this paper, all reported $UBVRI$ magnitudes are on the Johnson (Vega) system and all $ugriz$ magnitudes are on the AB system, unless otherwise specified.


	\begin{figure*}
	\begin{center}
	\begin{tabular}{c}
	\includegraphics[width=17.5cm]{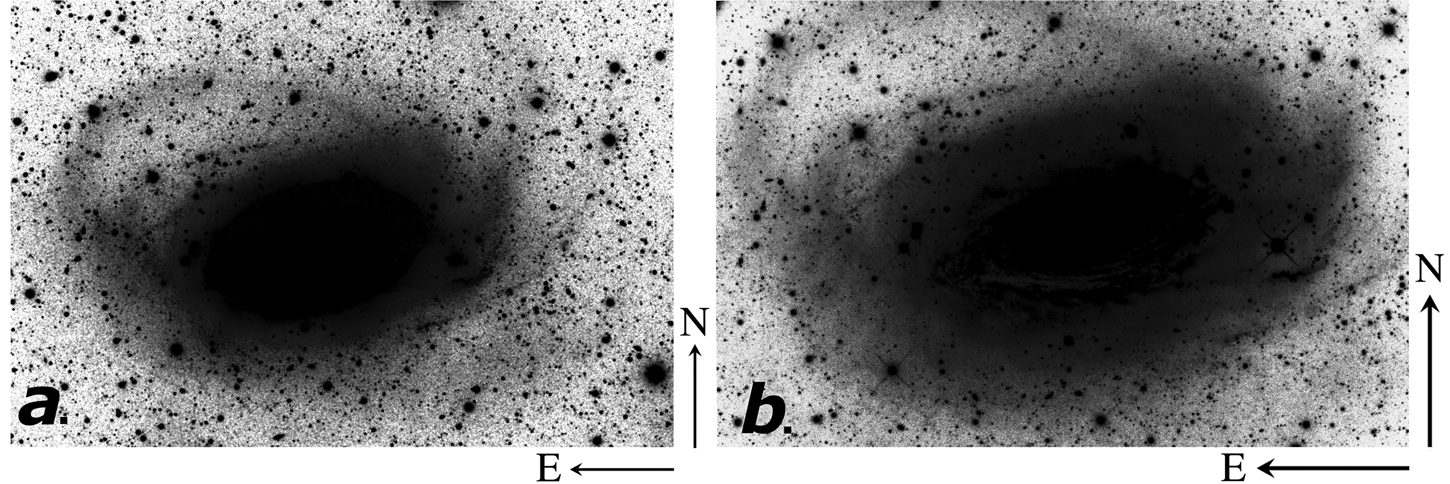}
	\end{tabular}
	\end{center}
	\caption[example] 
	{ \label{fig:GralakGabanyM63} 
	The M63 stellar tidal stream detection and confirmation images. The orientation and scale are indicated by the 6$\arcmin$ (12.6 kpc) long arrows in the lower-right corners of each image. Shown are the final 31.7 hour image obtained with the G-NMS 0.16 m ($a$) and the final 11.0 hour image obtained with the BBRO 0.5 m ($b$) resulting from the combination of the images listed in Table \ref{table:explog} after histogram equalization. The images have been cropped to 35.2$\arcmin$ $\times$ 25.7$\arcmin$ ($a$) and 26.9$\arcmin$ $\times$ 17.3$\arcmin$ ($b$).}
	\end{figure*} 
	
	\begin{deluxetable}{lcc}
	\tabletypesize{\scriptsize}
	\tablecaption{List of Data and Exposure Times\label{table:explog}}
	\tablehead{
		\colhead{Telescope} & \colhead{Filter} & \colhead{Tot. Exp. Time}\\
		& & \colhead{(minutes)}}
	\startdata
	G-NMS 0.16 m & CL & 790\\
 	& Red & 390\\
 	& Green & 390\\
 	& Blue & 330\\
 	& & \textbf{1900}$^\ast$\\ [1ex]
	BBRO 0.5 m & CL & 405\\
 	& Red & 90\\
 	& Green & 54\\
 	& Blue & 108\\
 	& & \textbf{657}$^\ast$\\ [1ex]
	MDO 0.8 m & $B$ & \textbf{390}$^\dagger$\\
 	& $R$ & \textbf{320}$^\dagger$\\
	\enddata
	\tablenotetext{$\ast$}{Total exposure time for images in Figure \ref{fig:GralakGabanyM63}.}
	\tablenotetext{$\dagger$}{Total exposure time for images in Figure \ref{fig:ChonisM63}.\\}
	\end{deluxetable}
	
\section{OBSERVATIONS AND DATA REDUCTION}\label{sec:Observations}
To obtain a clearer view of the barely detectable features first described by \citet{vanderKruit1979}, we obtained deep, wide-field images of M63 with a small 0.16 m telescope and first presented these data in \citet{martinezDelgado2010}. For completeness, these observations will be described below. We confirm the detection and study these features in more detail through new follow-up observations with two additional telescope-instrument combinations. Those less concerned with the technical observational details should skip directly to $\S$\ref{sec:LSBFeatures} for our findings.   
	
\subsection{Gralak-New Mexico Skies 0.16 m Telescope}\label{subsec:G-NMSobs}
As first presented in \citet{martinezDelgado2010}, low surface brightness features were detected in deep optical images of M63 taken with R. Gralak's 0.16 m Astro-Physics Starfire 160 $f$/5.9 Apochromatic refractor during dark-sky observing runs between April 2007 and February 2008 at New Mexico Skies Observatory situated in the Sacramento Mountains (New Mexico, USA - this telescope shall be referred to hereafter as G-NMS 0.16 m). The 0.16 m telescope is outfitted with a Santa Barbara Instruments Group (SBIG) STL-11000M\footnote{See \textit{http://www.sbig.com/sbwhtmls/large$\_$format$\_$cameras.htm\#STL-11000M} for the camera's specifications} CCD camera. This setup yields a large 131.2$\arcmin$ $\times$ 87.5$\arcmin$ field of view (FOV) at an image scale of 1.96$\arcsec$ pixel$^{-1}$. The data set consists of deep exposures (with individual exposure times of 20 or 30 minutes) taken through four non-standard photometric filters: a wide, non-infrared ``Clear-Luminance" filter (CL; $3500 < \lambda$ (\AA) $< 8500$) as well as Red, Green, and Blue filters from the SBIG-Custom Scientific filter set. Observations in these filters were intended for constructing a true-color image of M63. The details of the data acquisition and reduction (including standard methods for bias and dark subtraction as well as flat-fielding and image coaddition) are identical to those described in similar work by our group \citep{martinezDelgado2008, martinezDelgado2009}. As summarized in Table \ref{table:explog}, the total combined integration time of the data taken in all filters is 31.67 hours (1900 minutes). 

The final combined image contains a very large dynamic range since the tidal stream is much fainter than M63's disk. As described in \citet{martinezDelgado2008, martinezDelgado2009}, the final image has been histogram equalized by the application of an iterative, non-linear transfer function to optimize contrast and detail in the faintest parts of the image. The resulting image can be seen in Panel $a$ of Figure \ref{fig:GralakGabanyM63}. A complex, faint outer structure can be seen in unprecedented detail surrounding M63, including the large loop feature that was alluded to originally by \citet{vanderKruit1979}.

	\begin{figure*}
	\begin{center}
	\begin{tabular}{c}
	\includegraphics[width=17.5cm]{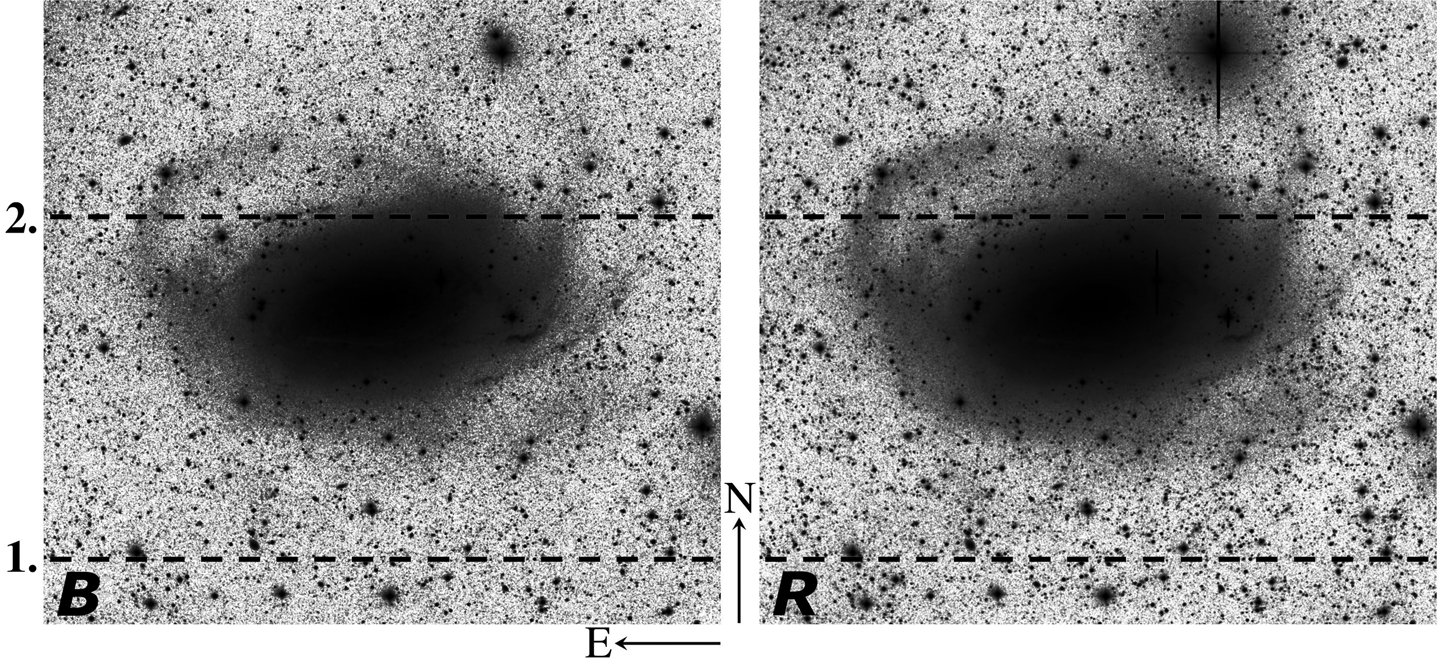}
	\end{tabular}
	\end{center}
	\caption[example] 
	{ \label{fig:ChonisM63} 
	MDO 0.8 m $B$ (\textit{left}) and $R$ (\textit{right}) final combined images. The orientation and scale are indicated by the 6$\arcmin$ (12.6 kpc) long arrows. The $B$ and $R$ images result from the combination of 6.5 and 5.3 hours of exposures, respectively. Each has been background subtracted as described in the text and histogram equalized. The field after trimming is 37.8$\arcmin$ $\times$ 34.9$\arcmin$. The dashed lines show the location of cuts through the image to demonstrate the effectiveness of the sky background subtraction, and are referred to in Figure \ref{fig:backgroundcuts}.}
	\end{figure*} 
		 
\subsection{BlackBird Remote Observatory 0.5m Telescope}\label{subsec:BBROobs}
We confirmed the above detection by re-examining a set of archived CCD images obtained during dark-sky observing runs in March and April of 2005 with the 0.5m $f$/8.3 Ritchey-Chr\'{e}tien telescope of the BlackBird Remote Observatory (BBRO), also located in the Sacramento Mountains. The data were acquired using an SBIG STL-11000M CCD camera, which yields a 29.8$\arcmin$ $\times$ 19.9$\arcmin$ FOV at an image scale of 0.45$\arcsec$ pixel$^{-1}$. As with the G-NMS data, the BBRO image set consists of multiple deep exposures (with individual exposure times of around 15 minutes) taken through the SBIG-Custom Scientific filter set. Data acquisition and standard data reduction procedures are also as described in the works cited above in $\S$\ref{subsec:G-NMSobs}. Table \ref{table:explog} summarizes the exposures included in 10.95 hours (657 minutes) of total integration. The resulting histogram equalized image is shown in Panel $b$ of Figure \ref{fig:GralakGabanyM63}. 
	
The BBRO 0.5m telescope has been used by our group in other studies of low surface brightness features around nearby spiral galaxies as a part of the pilot survey discussed in $\S$\ref{sec:Introduction} (\citealp{martinezDelgado2010}; see also \citealp{martinezDelgado2008,martinezDelgado2009}, \citealp{trujillo2009}, and \citealp{sollima2010}). From past experience (which will be confirmed later in this work), we estimate that we are able to detect faint features in the BBRO image (and the G-NMS final image) to a surface brightness of $\mu_{R}$ $\lesssim$ 27 mag arcsec$^{-2}$. 

\subsection{McDonald Observatory 0.8 m Telescope}\label{McDobs}
Since the above observations yield only limited photometric information, we have obtained follow-up observations using the McDonald Observatory (MDO) 0.8 m telescope. These observations will allow us to characterize portions of the faint, outer structure and potentially differentiate the tidal stream from other extended stellar halo and disk components.

The 0.8 m telescope utilizes a wide-field instrument, the Prime Focus Corrector (PFC), at $f$/3.0 \citep{claver1992}. This instrument contains a front-side illuminated Loral-Fairchild 2048$\times$2048 CCD with 15$\mu$ pixels that covers a 46.2$\arcmin$ $\times$ 46.2$\arcmin$ FOV at an image scale of 1.355$\arcsec$ pixel$^{-1}$. The detector has a read noise of 5.87 e$^{-}$ and a gain of 1.60 e$^{-}$ ADU$^{-1}$. The instrument covers a wide spectral range in the optical and near-IR from 3000 \AA\ to 1 $\mu$m, which is divided by a standard Bessel $UBVRI$ filter set. Deep images were acquired in $B$ and $R$ during a dark-sky observing run in April 2009 and are also summarized in Table \ref{table:explog}. All on-sky images were taken at an airmass $X < 1.6$ ($\sim70$\% of all images were acquired with $X < 1.2$) and were dithered to reduce systematics. The observing conditions were generally very good; however, two nights had relatively variable transparency conditions. 

\subsubsection{Standard Data Reduction}\label{subsubsec:reductions}
Standard data reduction procedures for overscan correction, bias subtraction, and dome flat-fielding were carried out using the CCDRED package in IRAF.\footnote{IRAF is distributed by the National Optical Astronomy Observatories, which is operated by the Association of Universities for Research in Astronomy, Inc., under cooperative agreement with the National Science Foundation.} No dark subtraction was applied as the dark signal is negligible in a 900 second exposure, the longest used in this study. Illumination correction is performed by taking \textit{at least} 3 blank-sky frames off-target (since M63 and its faint outer structure occupy a large fraction of the CCD's imaging area) per filter, interspersed evenly throughout each night at identical exposure times to the science frames. These sky images were median-combined with a $\sigma$-clipping algorithm to remove stars. This simple method removes all but the faintest parts of the outer stellar point spread functions (PSF) that happen to overlap between individual frames and evade rejection. The PSF residuals were found to be $\sim25\%$ of the large-scale sky variation, randomly distributed throughout the combined frame, and relatively small in extent (typically occupying $<1\%$ of the total image area, or less than a few $\times 10^4$ pixels). These resulting images were subsequently fit using a fifth order, two-dimensional Legendre polynomial. The fit for each filter's combined blank-sky frame serves as the illumination correction and was applied to each flat-fielded science image. Each calibrated science image was then examined for variable transparency through aperture photometry of several bright stars in the field. Those frames with instrumental magnitudes lying $>2\sigma$ from the mean were discarded. The remaining images were then average-combined. The final $B$ and $R$ images are combinations of 26$\times$900 second images (6.5 hours) and 32$\times$600 second images (5.3 hours), respectively. 

	\begin{deluxetable*}{lcccc}
	\tabletypesize{\scriptsize}
	\tablecaption{Random Noise Sources: MDO 0.8m - $R$ Image$^\ast$\label{table:ErrorBudget}}
	\tablehead{
	 	& \colhead{Per Single Pixel} & & \colhead{75$\times$75 Pixel Bin} & \\		
	 	Source of Uncertainty & \colhead{(ADU)} & \colhead{(\%)$^\#$} & \colhead{(ADU)} & \colhead{(\%)$^\#$}}
	\startdata
	Image Read Noise$^\dagger$ & 0.65 & 0.03 & 0.01 & 0.00 \\
	Image Photon Noise$^\dagger$ & 6.70 & 0.29 & 0.09 & 0.00 \\
	Flat-Field Photon Noise$^\ddagger$ & 4.42 & 0.19 & 0.06 & 0.00 \\
	Surface Brightness Fluctuation$^{\ast\ast}$ & 2.21 & 0.09 & 0.03 & 0.00 \\ [2ex]
	\textbf{Total Random Noise} & \textbf{8.33} & \textbf{0.36} & \textbf{0.11} & \textbf{0.01}\\
	\enddata
	\tablecomments{See $\S$\ref{subsubsec:errors} for details on the sources of uncertainty listed in this table.}
	\tablenotetext{$\ast$}{Average source ADU = 9.88; average sky ADU = 2287.52.}
	\tablenotetext{$\#$}{Percentage of the average total flux per pixel.}
	\tablenotetext{$\dagger$}{Calculated for 32 science images.}
	\tablenotetext{$\ddagger$}{Calculated for 25 flat-field frames with average 25000 ADU.}
	\tablenotetext{$\ast\ast$}{Calculated from \citet{tonry1988}, Equation 12.\\}
	\end{deluxetable*}
	
\subsubsection{Background Sky Modeling and Subtraction}\label{subsubsec:backgroundsub}
As will be seen, the components of the faint outer regions of M63 we are most interested in are $\sim$10 ADU above the background. Thus, careful background subtraction of the final images must be obtained for useful surface photometry. Subtraction of a constant value for the sky background is complicated by the fact that the night sky is not uniform over such a large FOV \citep{wild1997}. Additionally, Figure \ref{fig:ChonisM63} shows that M63 and its faint outer regions occupy a large fraction of the imaged area. With this in mind, a simple mask and global modeling technique (using a 2-D analytical function, for example) is likely not the best choice for background determination. \citet{zheng1999} found that such a method systematically under or over-fits certain regions of their images used for deep surface photometry. Instead, they fit the sky piece-wise in a row-by-row and column-by-column fashion using low-order polynomials. We adopt this method of sky subtraction and briefly describe it here as it applies to our observations. 

We begin by masking all sources in the $B$ and $R$ final images so that the wings of stellar PSFs, small galaxies, and all faint regions of M63 were covered to $\lesssim$15$\%$ of the sky variation. This level was chosen because there is a trade-off between the surface brightness one can mask the PSF wings to and leaving enough pixels after masking to properly sample the fit \citep{zheng1999}. For both the $B$ and $R$ images, $\sim$51\% of pixels were masked leaving $\sim$49\% available for the fitting process. We then twice model the background piece-wise using third order polynomials: once fitting each row and again fitting each column, excluding masked pixels and those lying $\pm$2$\sigma$ from the mean value of all non-masked pixels in the row or column. The models fit from rows and from columns are then averaged to produce a final background model that is better sampled, and was finally smoothed using a large Gaussian kernel ($\sigma$=80 pixels; 4$\sigma$ cutoff) to eliminate any small artifacts from the modeling process. The result is then subtracted from the final $B$ and $R$ images as derived for each individually. The accuracy of the modeling and subtraction will be evaluated below in $\S$\ref{subsubsec:errors}. The resulting sky models show that the final $B$ and $R$ images have backgrounds that are already quite free of variation. The fit maximum-to-minimum variation from average is 0.74\% for $B$ and 0.73\% for $R$.   
	              
\subsubsection{Photometric Calibration}\label{subsubsec:zeropt}
M63 lies in a Sloan Digital Sky Survey (SDSS; \citealp{york2000}) field. Thanks to the PFC's wide FOV, there are a large number of isolated, dim stars (i.e., not saturated in SDSS or our images) available by which we can obtain a photometric calibration for $B$ and $R$ using ``tertiary" $gri$ SDSS standards following the method and transformations of \citet{chonisGaskell2008}. Their transformations (Equations 1 and 3 in that work) are valid only over the ranges $0.2<g - r<1.4$ and $0.08<r - i<0.5$. To determine that these color ranges are appropriate for our observations, we plot $B-R$ vs. $g-r$ and $B-R$ vs. $r-i$ color-color diagrams using the large sample of matched \citet{stetson2000} $BVRI$ standard photometry and SDSS DR4 \citep{adelmanmccarthy2006} $ugriz$ photometry ($>1200$ stars) from \citet{jordi2006}. Using these diagrams, we verify the linearity between $B-R$ and the SDSS colors over the limited color ranges. Given that the range of $B-R$ galaxy colors ($0.4<B - R<1.8$) from the large, morphologically diverse sample of \citet{jansen2000} comfortably fits within these linear regions, we expect the transformations of \citet{chonisGaskell2008} to be adequate for the range of $B-R$ colors we might expect to observe in the M63 field.

Instrumental magnitudes of 113 stars in each $B$ and $R$ image were obtained; SDSS photometry for these stars was obtained from the DR7 SkyServer\footnote{DR7 SkyServer: \textit{http://cas.sdss.org/dr7/en/}} \citep{abazajian2009}. Following the selection criteria of \citet{chonisGaskell2008}, 63 of these stars were transformed from SDSS magnitudes to $B$ and $R$ and compared to the instrumental magnitudes. The residuals were fit with a constant zero-point offset as well as a small color term in the instrumental $B-R$ color. The standard error in the mean of the residuals after this correction is given by $\sigma_{phot,x}$(63)$^{-1/2}$, where $\sigma_{phot,x}$ is the residuals' standard deviation in photometric band $x$ (where $x$ is either $B$ or $R$). This results in an uncertainty of 0.010 mag and 0.007 mag for $B$ and $R$ respectively. This uncertainty's contribution to the total error budget will be discussed below.

Note also that all reported magnitudes for M63 have been additionally corrected for Galactic extinction, but not for galaxy inclination. From \citet{schlegel1998}, $A_{B}$ = 0.076 and $A_{R}$ = 0.047; according to the NASA/IPAC Infrared Science Archive,\footnote{NASA/IPAC Infrared Science Archive: \textit{http://irsa.ipac.caltech.edu/applications/DUST}} $A_{V}$ varies over the MDO 0.8 m FOV by no more than $\pm 0.01$ mag in this region of the sky.

	\begin{figure*}
	\begin{center}
	\begin{tabular}{c}
	\includegraphics[width=17.5cm]{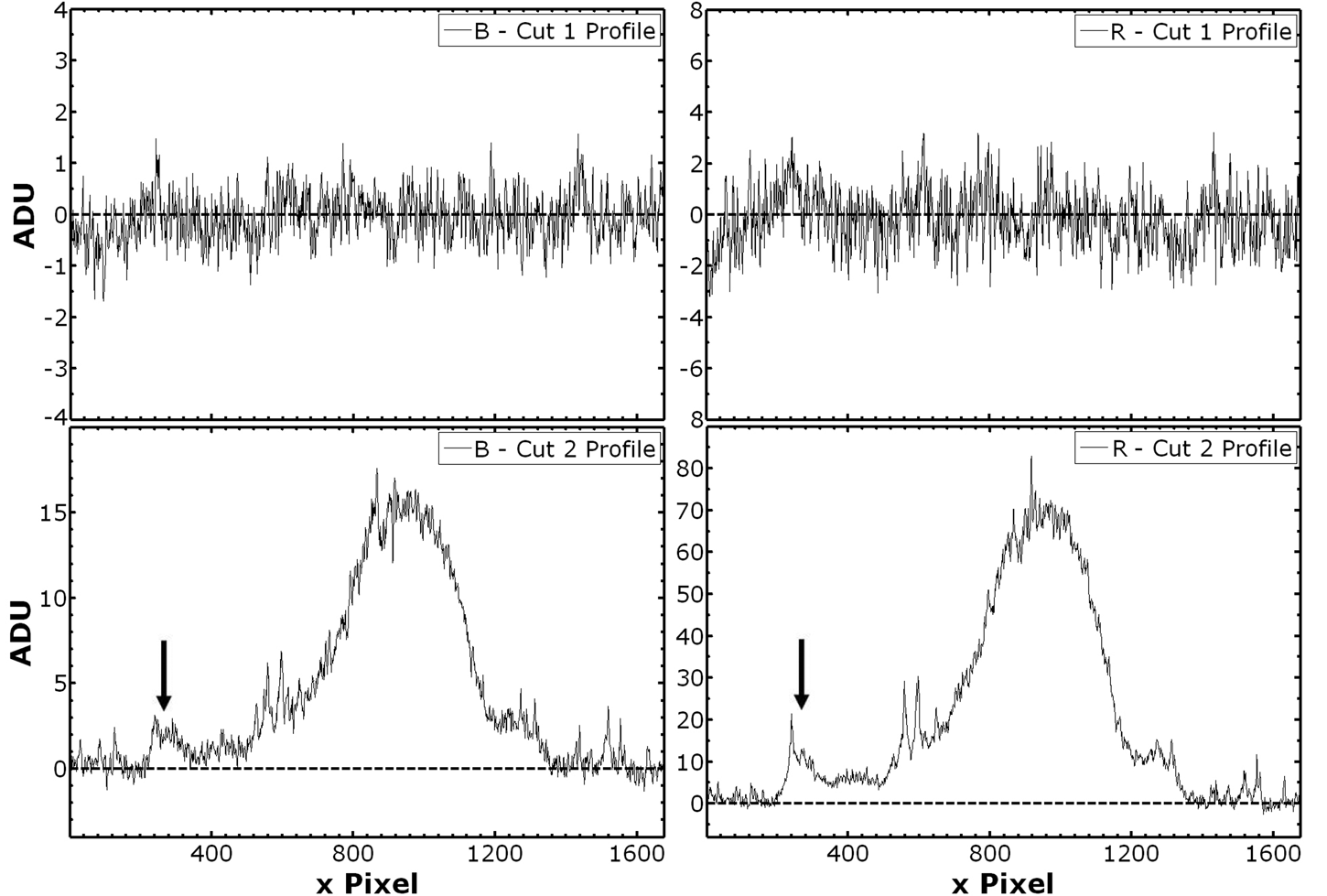}
	\end{tabular}
	\end{center}
	\caption[example] 
	{ \label{fig:backgroundcuts} 
	Profiles showing the effectiveness of the sky background subtraction process. The location of the cuts can be seen in Figure \ref{fig:ChonisM63}. Each profile is median-combined over a 30 pixel-wide strip. Cut 1 (top panels) is through a region with relatively few sources (which are masked) to show typical residuals of the background subtraction.  The background of the $B$ ($R$) image has a typical value before subtraction of 364 (2283) ADU. Cut 2 (bottom panels) is through the stellar stream and the outer stellar halo of the galaxy with no sources masked to show the relative signal strength of the photometric features of interest as compared to the subtracted background. The arrows indicate the location of signal due to the stellar stream.}
	\end{figure*} 
		
\subsubsection{Estimation of Measurement Uncertainties}\label{subsubsec:errors}
We discuss the various sources of uncertainty in our surface brightness measurements and describe the methods by which we estimate them. The sources of uncertainty can be divided into two main categories: 
	
\textit{1. Random Noise} - Sources of random noise, such as read noise from the CCD read-out process and Poisson noise due to photon counting for both science frames and flat-field frames, are calculated using standard methods. An additional source of random noise is surface brightness fluctuations from the measured photometric feature itself. Because the faint substructures of interest are composed of individual stars, each pixel contains noise from counting statistics (because there are an average, finite number of stars contained in the solid angle on the sky subtended by an individual pixel). The noise contributed by this surface brightness fluctuation is given by the square-root of Equation 10 in \citet{tonry1988}. In their equation, we use $d$ = 7.2 Mpc; for $B$ ($R$), $t$ = 900 (600) seconds, the magnitude corresponding to 1 ADU/s $m_{1}$ = 20.91 (21.51) mag, and $\bar{M}$ = 1 (0, to be conservative). Finally, $\bar{g}$ is the average signal (in ADU) of the area of the galaxy being measured. Sampling more pixels in a given measurement bin reduces random noise (e.g., as $n^{-1/2}$, where $n$ is the number of pixels in the bin). Thus, we can reduce these noise sources to near negligible levels, as will be the case for all signal levels considered in this work, by sampling many pixels covering a photometric feature. To illustrate this and to give a sense of the level of random noise in our images, we sample a 75$\times$75 pixel box centered in a region of the stellar stream in the $R$ band image containing on average 9.88 ADU pixel$^{-1}$ from the stream and 2287.52 ADU pixel$^{-1}$ from the sky. Based on the photometric calibration discussed in $\S$\ref{subsubsec:zeropt}, this results in a surface brightness for the stellar stream of $\mu_{R}$ = 26.0 mag arcsec$^{-2}$. The noise in this bin is compared to that in a single pixel and is tabulated in Table \ref{table:ErrorBudget}. The noise levels are reported both in ADU and as a percentage of the average total flux per pixel.
	
\begin{figure}
	\begin{center}
	\begin{tabular}{c}
	\includegraphics[width=8.1cm]{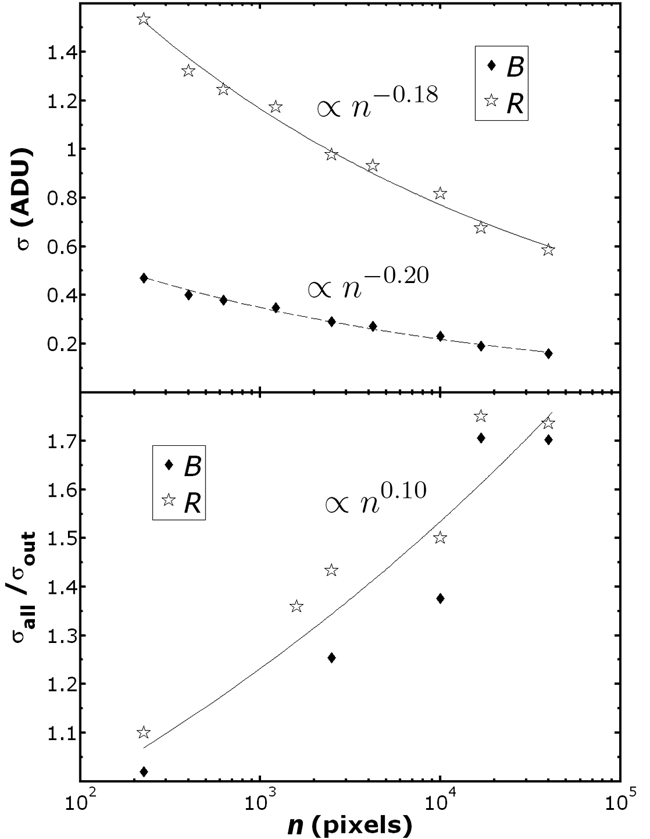}
	\end{tabular}
	\end{center}
	\caption[example] 
	{ \label{fig:stddevscale} 
	The standard deviation $\sigma$ (\textit{top}) of the distribution of the median of pixels contained in various sized bins placed in the non-masked portions of the $B$ (diamonds) and $R$ (stars) images after sky subtraction as a function of the total number of pixels $n$ contained in each bin. The thin solid lines are power law fits for each set of data points. The same is also shown for the quantity $\sigma_{all}/\sigma_{out}$ (\textit{bottom}), which is the ratio of the standard deviation of the distribution of the median of pixels contained in bins placed over the entire combined blank-sky illumination correction image after background modeling and subtraction to that placed outside the dummy mask.}
	\end{figure}

\textit{2. Non-Random Errors} - The main source of non-random error in photometry of large, low surface brightness objects results from the inevitably imperfect sky subtraction process. Figure \ref{fig:backgroundcuts} shows profiles through the $B$ and $R$ images whose locations are indicated by the dashed lines in Figure \ref{fig:ChonisM63}. Cut 1 shows that the mean of the background after subtraction is very close to 0 and that there are only slight indications of systematic errors introduced by the sky subtraction process, which are small in amplitude and extent. Cut 2 shows the signal from the stellar stream and faint disk light of M63 after background subtraction. For a more quantitative look at the background after sky subtraction, we calculate the median of pixels contained in various sized bins placed in a grid across the masked, background subtracted final images in which we consider only bins with $>$85$\%$ unmasked pixels. On all scales, the distributions of bin medians are nearly Gaussian in shape. As noted in \citet{zheng1999}, these distributions are slightly skewed in a positive sense given that the faint, extreme outer parts of bright PSFs are not necessarily masked since masking them entirely would leave very few pixels with which to fit the background. This slightly increases the distributions' standard deviations. In the presence of the positive skew, we note that the modes of the distributions are 0, indicating that the backgrounds were properly subtracted. The top panel of Figure \ref{fig:stddevscale} shows the trend of the distributions' standard deviations $\sigma$ with the number of pixels $n$ contained in the bin for both $B$ and $R$ images. Note that $\sigma$ measures \textit{both} random noise in the background and the uncertainty due to sky subtraction. By inspection of the top panel of Figure \ref{fig:stddevscale}, one can see that $\sigma$ is much larger than would be indicated by random noise alone (see Table \ref{table:ErrorBudget}). This excess uncertainty in the measured values of $\sigma$ is due to the systematic error in the background modeling and subtraction. As such, the data are fit well with power laws in $n$ having exponents of -0.20 and -0.18 for $B$ and $R$, respectively, rather than -0.50 as would be expected from only random noise.  

The pixels sampled above can only be those from the sky that were unmasked and directly fit. Thus it is unknown to what degree the sky fitting method properly interpolates the sky values under the mask where measurements will later be taken. To quantify this, we combine all frames taken off-target for illumination correction for each filter and median-combine them with a $\sigma$-clipping algorithm, removing the signature of field stars as described in $\S$\ref{subsubsec:reductions}. The mask used for modeling the background of the science frames was then applied as a dummy mask. We use the remaining pixels to model the blank-sky background, using third order polynomials as in $\S$\ref{subsubsec:backgroundsub}. The final fit was subtracted from the combined blank-sky frames and the distribution of bin median values was determined outside the dummy mask, as described in the preceding paragraph. We denote this distribution's standard deviation as $\sigma_{out}$. We also sample bins over the entire image with the dummy mask removed after subtracting the sky model since there is no diffuse source under it. The standard deviation of this distribution is also calculated and is denoted $\sigma_{all}$. We find for both $B$ and $R$ that in general, $\sigma_{all}/\sigma_{out} > 1$, which is indicative of a less than ideal interpolation in the masked region where the galaxy is located and the sky is not directly fit. This quantity is not affected by the PSF residuals in the $\sigma$-clipped median combine since their effect cancels when taking the ratio due to the random distribution throughout the image. Finding that $\sigma_{all}/\sigma_{out} > 1$ is not surprising given that M63 occupies a large fraction of the total image area. In the bottom panel of Figure \ref{fig:stddevscale}, we show the trend of $\sigma_{all}/\sigma_{out}$ with $n$ for both $B$ and $R$ images. Both data sets are well fit together with a single power law in $n$ with an exponent of 0.10. This analysis with that described in the preceding paragraph shows that the measurement uncertainty after background subtraction has only a weak dependence on the size of the feature being measured. It also shows that the background subtraction accuracy is limited by the faint, outer portions of bright PSFs and the inability to directly model the sky under the large galaxy mask.

The remaining source of uncertainty is in the photometric zero-point determined in $\S$\ref{subsubsec:zeropt}. Since this uncertainty is in a systematic photometric offset, it is simply added into the total error budget after its conversion to magnitudes. 
	
\textit{Sample Error Estimate} - Using the above empirical estimates of the measurement uncertainty after subtracting the modeled background, we give a sample error estimate for the $75\times75$ pixel box centered on the tidal stream in which we estimated the contribution of uncertainty due to random noise for the $R$ image. We first estimate $\sigma$ (top panel of Figure \ref{fig:stddevscale}) and multiply this value by the corresponding scaling factor (bottom panel) to take into account the imperfect interpolation under the mask where the stream is located. For a $75\times75$ pixel box, this corresponds to an uncertainty of 1.21 ADU. We then add the small contribution from random noise (here, 0.11 ADU from Table \ref{table:ErrorBudget}). After conversion to magnitudes and addition of the uncertainty in the photometric zero-point, this yields a surface brightness of $\mu_{R} = 26.0^{+0.2}_{-0.1}$ mag arcsec$^{-2}$.

	\begin{figure*}
	\begin{center}
	\begin{tabular}{c}
	\includegraphics[width=17.5cm]{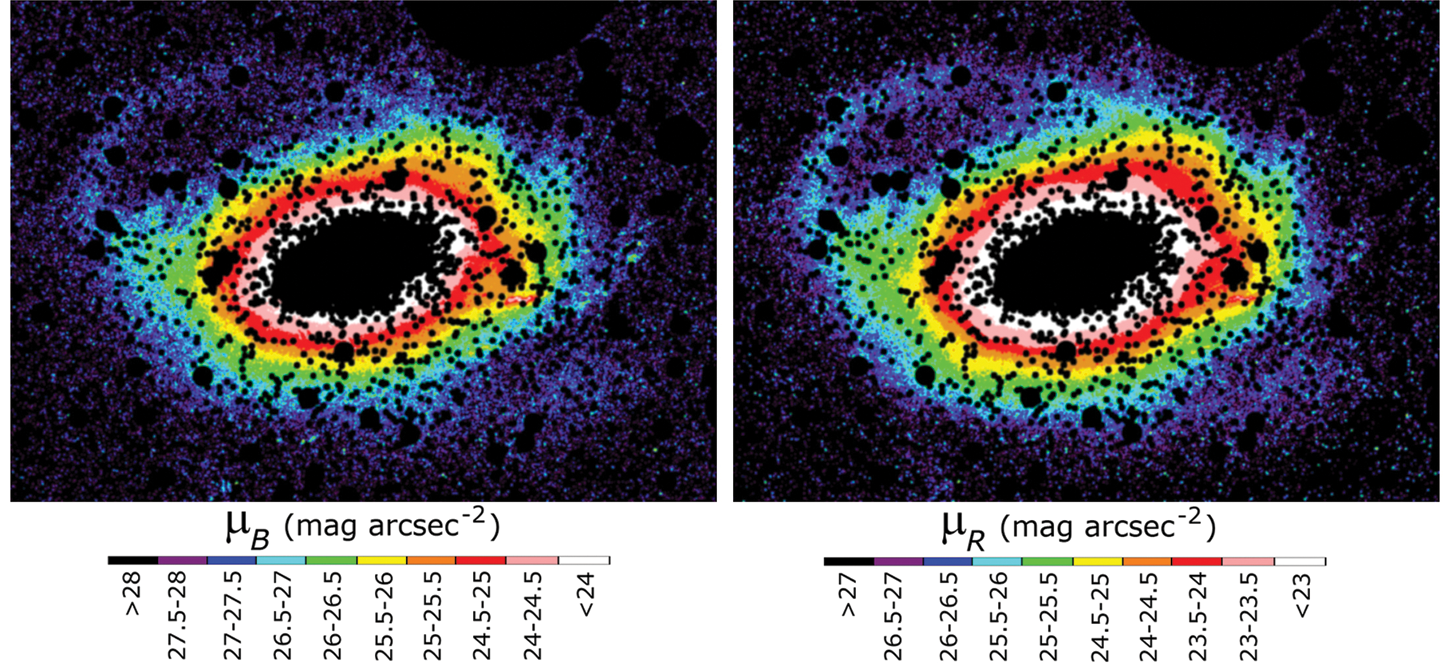}
	\end{tabular}
	\end{center}
	\caption[example] 
	{ \label{fig:SBmap} 
	$B$ (\textit{left}) and $R$ (\textit{right}) surface brightness contour maps, derived based on the description given in $\S$\ref{subsubsec:maps}. Note that the contour scales for each are not identical.}
	\end{figure*} 
	
	\begin{figure}
	\begin{center}
	\begin{tabular}{c}
	\includegraphics[width=8.1cm]{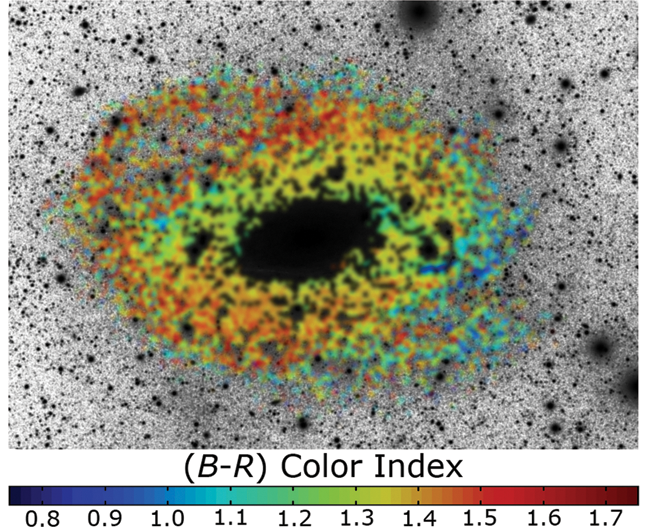}
	\end{tabular}
	\end{center}
	\caption[example] 
	{ \label{fig:Colormap} 
	$B-R$ color index map, derived based on the description given in $\S$\ref{subsubsec:maps} and overlaid on the G-NMS image. Typical errors in $B-R$ are $\pm0.2$ mag for a feature with $\mu_{R} \approx 26$ mag arcsec$^{-2}$ (refer to Figure \ref{fig:SBmap} for the surface brightness of various features).}
	\end{figure} 
	
\subsubsection{Surface Brightness Maps}\label{subsubsec:maps}
We create a new mask covering the bright disk of M63 and all intervening field stars while leaving the faint outer features exposed. The remaining unmasked pixels are converted from ADUs to magnitudes, producing $B$ and $R$ surface brightness maps of the faint structures based on the images shown in Figure \ref{fig:ChonisM63}. These maps are shown in Figure \ref{fig:SBmap}. The brightest isophote outside of the mask is $\mu_{R}$ $\approx$ 23 mag arcsec$^{-2}$. As can be seen, we can reliably measure features to $\mu_{R}$ $\approx$ 27 mag arcsec$^{-2}$. 

From these maps, we have created a $B-R$ color index map. Since colors are inherently more uncertain than an individual surface brightness measurement, we bin the $B-R$ array in 7$\times$7 pixel boxes, taking the median of the pixels in the bin while ignoring those masked. Only bins with $>$60\% unmasked pixels are considered. The resulting color index map can be seen overlaid on the G-NMS image in Figure \ref{fig:Colormap}. We find that $B-R$ colors in M63's faint outer structures range from $\sim$0.8 to $\sim$1.7, with noticeable systematic color gradients present. The implication of both the surface brightness maps and the $B-R$ color map are discussed in the following section. 

\section{LOW SURFACE BRIGHTNESS FEATURES}\label{sec:LSBFeatures}
\begin{figure*}
	\begin{center}
	\begin{tabular}{c}
	\includegraphics[width=17.5cm]{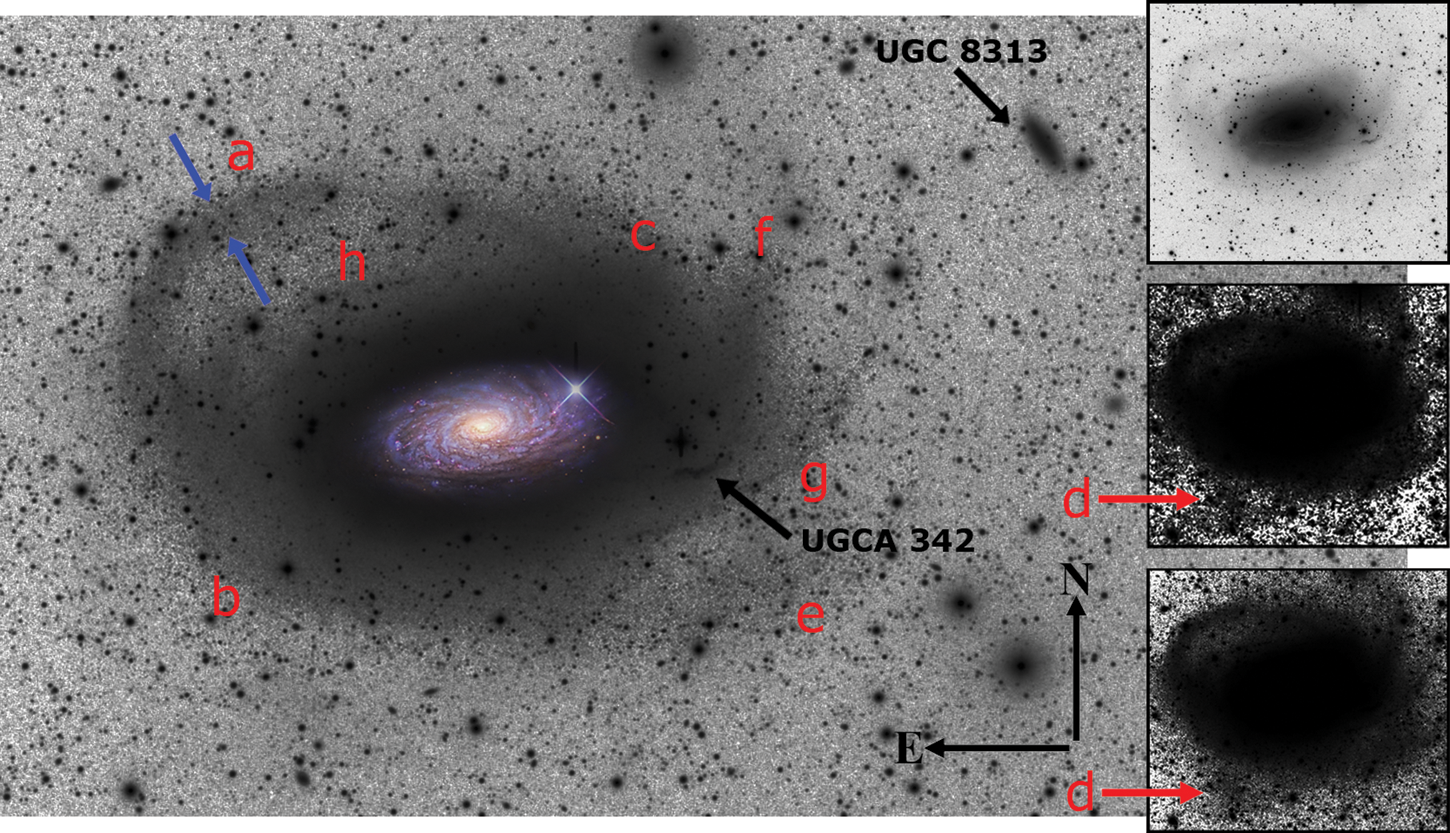}
	\end{tabular}
	\end{center}
	\caption[example] 
	{ \label{fig:photofeatures} 
	Photometric features measured by the MDO 0.8 m are labeled $a$ - $h$ on the G-NMS image. For reference, a color image of M63's disk constructed from the BBRO data has been superimposed, indicating the more familiar extent of this galaxy. The blue arrows indicate the location where the width of the main loop was measured. The red arrows in the insets indicate the location of an extremely dim feature ($d$). The middle inset is the MDO $R$ band image while the lower is confirmation from the G-NMS image. For reference, we show in the top inset a less dramatic stretch of the G-NMS image to more clearly show the inner regions of the stellar halo. Note that that each photometric feature is explicitly outlined in Figure \ref{fig:photofeaturesextent}.}
	\end{figure*}

\begin{figure}
	\begin{center}
	\begin{tabular}{c}
	\includegraphics[width=8.1cm]{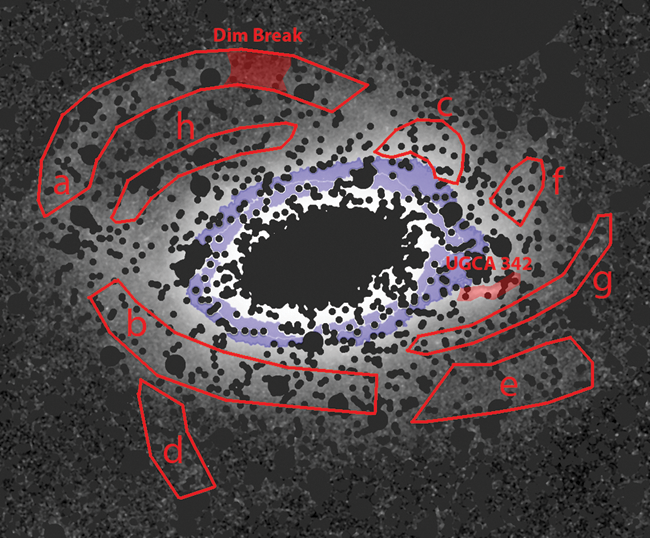}
	\end{tabular}
	\end{center}
	\caption[example] 
	{ \label{fig:photofeaturesextent} 
	MDO $R$ band image shown with foreground stars, background galaxies, and the disk of M63 masked and leaving the faint outer disk and stellar halo substructure exposed for measurement. Each photometric feature labeled $a$ - $h$ in Figure \ref{fig:photofeatures} has been enclosed in red, showing the area of pixels used for the average surface brightness measurements in Table \ref{table:PhotoData}. The measured areas of UGCA 342 and the dim break in the tidal stream are shaded red while the measured area around the $\mu_{R}$ = 25 mag arcsec$^{-2}$ isophote is shaded blue.}
	\end{figure}

Based on our deep imaging, we have identified 8 distinct photometric features of interest, the most prominent of which is the giant loop structure first discovered by \citet{vanderKruit1979} to the north-east of the galaxy's disk resembling a flower petal (which is fitting, considering M63's popular name). We identify and label these features in Figure \ref{fig:photofeatures}. For all features, we have calculated $B$ and $R$ surface brightness and $B-R$ color indices (averaged over the entire spatial extent of each feature, as illustrated in Figure \ref{fig:photofeaturesextent}). The measurements are presented in Table \ref{table:PhotoData}. 

	\begin{deluxetable*}{clllll}
	\tabletypesize{\scriptsize}
	\tablecaption{Measured Properties of Low Surface Brightness Features in Messier 63\label{table:PhotoData}}
	\tablehead{
 	& & \colhead{$\mu_{B}$} & \colhead{$\mu_{R}$} & \colhead{$B$ - $R$} & \\		
 	Feature & & \colhead{(mag arcsec$^{-2}$)} & \colhead{(mag arcsec$^{-2}$)} & \colhead{(mag)} & \colhead{Association/Comment}}
	\startdata
 	\textit{a} & Loop & 27.6$\pm$0.2 & 26.1$\pm$0.1 & 1.5$\pm$0.2 & Stellar Stream \\ [1.0ex]
 	& Dim Break & 28.3$^{+0.7}_{-0.4}$ & 26.8$\pm$0.3 & --- & \\ [2.0ex]
 	\textit{b} & . . . . . . . . . . . . . . . . . & 27.0$\pm$0.1 & 25.6$\pm$0.08 & 1.5$\pm$0.1 & Tidal Debris \\ [2.0ex]
 	\textit{c} & . . . . . . . . . . . . . . . . . & 25.80$^{+0.05}_{-0.04}$ & 24.48$\pm$0.03 & 1.32$^{+0.06}_{-0.05}$ & Tidal Debris \\ [2.0ex]
 	\textit{d} & . . . . . . . . . . . . . . . . . & 28.8$^{+1.0}_{-0.5}$ & 27.4$^{+0.7}_{-0.4}$ & --- & Stellar Stream \\ [2.0ex]
 	\textit{e} & . . . . . . . . . . . . . . . . . & 27.9$\pm$0.3 & 26.7$\pm$0.2 & 1.2$\pm$0.4 & Tidal Debris/Spiral Association \\ [2.0ex]
 	\textit{f} & . . . . . . . . . . . . . . . . . & 26.46$^{+0.10}_{-0.09}$ & 25.20$\pm$0.07 & 1.3$\pm$0.1 & Tidal Debris/Spiral Association \\ [2.0ex]
 	\textit{g} & Small Clumps & 26.9$\pm$0.1 & 26.0$\pm$0.1 & 1.0$\pm$0.1 & UV-Extended Disk \\ [1.0ex] 
 	& UGCA 342 & 24.83$\pm$0.03 & 23.96$\pm$0.03 & 0.87$\pm$0.04 & \\ [2.0ex]
 	\textit{h} & Small Clumps & 26.33$\pm$0.09 & 25.39$\pm$0.09 & 0.9$\pm$0.1 & UV-Extended Disk \\ [2.0ex]
 	M63 & Outer Disk/Stellar Halo & 24.83$\pm$0.02 & 23.52$\pm$0.01 & 1.31$\pm$0.02 & $\mu_{R}$ = 23.5 Isophote
	\enddata
	\end{deluxetable*}
	
\subsection{The Tidal Stream of M63}\label{subsec:tidalstream}
In Figure \ref{fig:photofeatures}, feature $a$ is the faint loop structure that apparently emerges from the East side of M63's disk and sweeps almost 180$^{\circ}$ around the system to the North-East with its center reaching 14.0$\arcmin$ ($\sim$29 kpc projected) from the galaxy's center and entering again at the North-West side. When first discovered by \citet{vanderKruit1979}, there was little doubt that this feature was real. However, it was unclear if it originated in M63 itself, or if it was instead ``high latitude reflection nebulosity in our Galaxy" (i.e., Galactic cirrus), analogous to that contaminating the field of M81 and M82 \citep{sollima2010}. Our deep images show that this feature is part of M63 based on similar morphology to the great-circle type tidal streams, arising from the recent merger and disruption of a lower mass satellite galaxy in a nearly circular orbit (\citealp{johnston2008, martinezDelgado2010}; see $\S$\ref{sec:Discussion} below for further discussion). This morphology is also reported for the stellar stream associated with the Sagittarius dwarf (e.g., \citealp{ibata2001b, martinezDelgado2004, law2010}) and around other near-by systems, such as NGC 5907 \citep{shang1998, martinezDelgado2008} and M83 \citep{malin1997}. 

To quantify the morphological characteristics of feature $a$, we have measured its projected orientation relative to the spiral disk of M63 by fitting ellipses to the MDO images. The ellipticity of M63's disk, measured at the $\mu_{B}$ = 25.0 mag arcsec$^{-2}$ isophote, is found to be $\epsilon_{M63}$ = 0.80. The flattening is thus $E_{M63}$ = 0.40, yielding an inclination of $i_{M63}$ = $\arccos(1 - E_{M63})$ $\approx$ 53.3$^{\circ}$. To check our ellipse fitting procedure, we verify that this value is in good agreement with that reported by the HYPERLEDA database ($i_{M63}$ $\approx$ 56.0$^{\circ}$). The stream's light distribution was traced and fit with an ellipse having a semi-major axis $a$ of 8.95\arcmin\ (18.7 kpc) and a minor-to-major axis ratio $q = 0.57$. We find that the position of the major-axis of the stream is tilted with respect to the major-axis of M63's disk, with a measured difference in position angle $\Delta\phi$ = 14.8$^{\circ}$. Note, however, that with no dynamical information at hand for the stream, we cannot use the ellipse fitting to yield its orientation on the sky plane due to the degeneracy with intrinsic orbital ellipticity.

We have measured the width of the stream, defined as the Full Width at Half Maximum (FWHM) of a Gaussian fit as in \citet{martinezDelgado2008}. Visual inspection of the stream shows that the width has little variability along its path. However, it is difficult to verify this at the locations where the stream appears to enter the galaxy's outer regions as the faint light and other features at high galactocentric radius contaminate our view of the stream. The stream width was measured from an extracted 200 pixel wide strip in the MDO $R$ band image that is perpendicular to the stream at the location indicated in Figure \ref{fig:photofeatures}. The strip was collapsed with a median-combine resulting in a single profile through the stream. Since the background is slightly brighter on the side of the stream that is closer to the galaxy, simple measures of FWHM would be biased toward larger values. To avoid this, we sample the background faint light from the galaxy in 100 pixel long zones on either side of the stream. Since the sampled strip is relatively small in length compared to the size of the galaxy's diffuse outer light, we approximate the galaxy contribution to the stream's light profile as linear and subtract it; the resulting profile is fit well with a Gaussian function. We estimate the FWHM of the stream to be $\sim$95\arcsec, which translates to 3.3 kpc at the 7.2 Mpc distance to M63. To verify that the $R$ band measurement is consistent with stream FWHM values reported in previous work from images constructed with SBIG-Custom Scientific filters (e.g., \citealp{martinezDelgado2008, martinezDelgado2009}), we have also measured the stream width using the G-NMS and BBRO images. To within 0.1 kpc, both measurements agree with that found above.   

The average surface brightness of the loop, as measured along its length within the FWHM measured in the preceding paragraph (see Figure \ref{fig:photofeaturesextent}), is $\mu_{R}$ = 26.1$\pm$0.1 mag arcsec$^{-2}$ with an average color index of $B-R$ = 1.5$\pm$0.2. The stream appears to be fairly constant in surface brightness and color along its length, except for an interesting small dim break in the stream's light distribution to the north, northeast of the galaxy (see Figure \ref{fig:photofeaturesextent}) lasting for $\sim$2.7\arcmin\ ($\sim$5.7 kpc in projection). The surface brightness in this dim break is very uncertain, but it is probably close to 1 mag arcsec$^{-2}$ dimmer than the rest of the stream. Possible causes of this surface brightness discontinuity in the otherwise coherent arc of the stellar stream are discussed below in $\S$\ref{subsec:nLoops}.

As was found in \citet{martinezDelgado2009} for the tidal stream of NGC 4013, the $B-R$ color index of this stream is quite consistent with the red colors of S0 galaxies from the sample of \citet{barway2005}. This color is also consistent with the redder dwarf galaxies in the Local Group \citep{mateo1998} and possibly suggests that the stream is dominated by an old ($\gtrsim$10 Gyr) stellar population, which is also predicted in general by the models of \citet{cooper2010}. Within 1$\sigma$, the measured stream color is similar to the faint outer regions of the disk and stellar halo of M63 (as measured at the $\mu_{R}$ = 23.5 mag arcsec$^{-2}$ isophote; see Table \ref{table:PhotoData}). However, the $B-R$ color index map (Figure \ref{fig:Colormap}) shows a slight systematic color gradient towards redder colors at larger disk radii, most notably near the outer eastern edges. This color gradient was also reported in the photographic study of \citet{vanderKruit1979}.   
	
	\begin{figure}
	\begin{center}
	\begin{tabular}{c}
	\includegraphics[width=8.1cm]{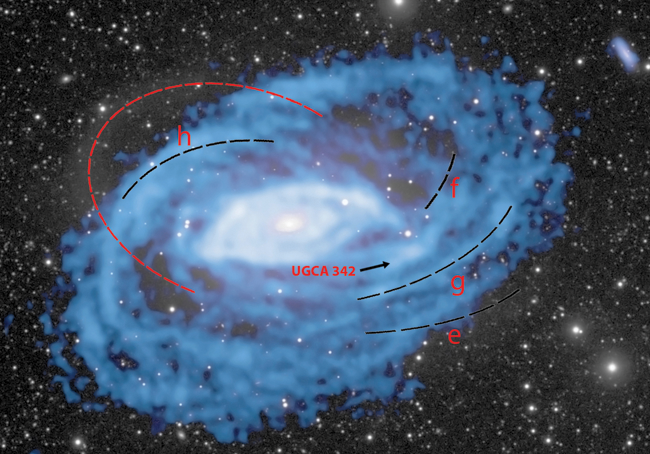}
	\end{tabular}
	\end{center}
	\caption[example] 
	{ \label{fig:HImap} 
	A comparison of an optical image (shown here in grayscale from the G-NMS 0.16 m telescope) with the distribution of HI gas around M63. The HI map from \citet{battaglia2006} is at 67$\arcsec$ resolution, has a column density detection limit of 0.10 M$_{\odot}$ pc$^{-2}$, and is shown in blue overlaid on the optical image. The red dashed curve represents the location of the stream ellipse fit described in $\S$\ref{subsec:tidalstream} and relevant photometric features are labeled.}
	\end{figure} 
	
Figure \ref{fig:HImap} shows a comparison of one of our optical images with an HI map at 67$\arcsec$ resolution from \citet{battaglia2006}. The tidal stream extends well beyond the significant HI emission of the disk and apparently has no gaseous component at the detection limit of this HI survey (0.10 M$_{\odot}$ pc$^{-2}$). Visible in this map is the significant HI warp of $\sim$20$^{\circ}$ as well as extended spiral structure. Some of the other very faint structures visible in our deep images are clearly recognizable in this map. For example, the conspicuous narrow trail of gas visible in the northeast edge of the HI disk corresponds to the optical feature labeled $h$ in Figure \ref{fig:photofeatures}. Additionally, the long line of gas on the opposite side of the HI disk from $h$ corresponds to optical feature $g$ with the large clump just inside of it corresponding to the over-density carrying the designation UGCA 342. As will be discussed in the following section, $g$, $h$, and UGCA 342 are likely not directly associated with the tidal stream. Finally, it is interesting to point out the enormous ``holes'' in the HI distribution at the locations where the tidal stream intersects the disk plane (see the ellipse fit tracing the stellar stream's light distribution in Figure \ref{fig:HImap}). Further discussion on the possibility of these voids being due to the passage of the disrupted dwarf through the HI disk is given in $\S$\ref{subsec:ParentEffect}.

\subsection{Other Features}\label{subsec:otherfeatures}

	\begin{figure}
	\begin{center}
	\begin{tabular}{c}
	\includegraphics[width=8.1cm]{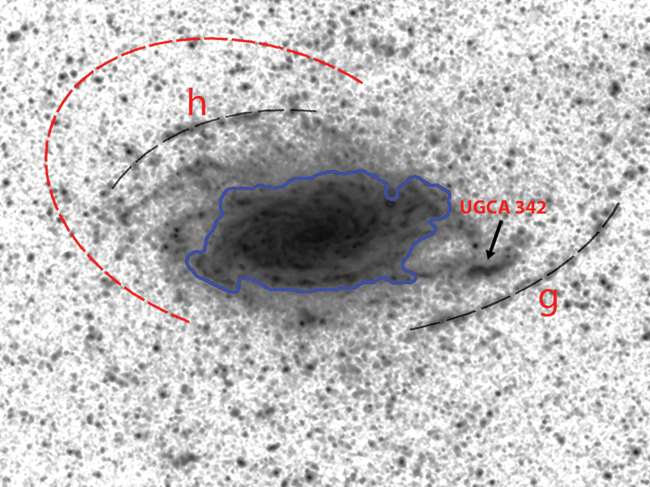}
	\end{tabular}
	\end{center}
	\caption[example] 
	{ \label{fig:GALEXmap} 
	An archival GALEX UV image ($1350 < \lambda$ (\AA) $< 2800$) showing M63's XUV disk, parts of which correspond to detected optical features $g$ and $h$. The blue contour is from \citet{thilker2007} and indicates the $\mu_{FUV}$ = 27.25 AB mag arcsec$^{-2}$ isophote. The red dashed curve represents the location of the stream ellipse fit described in $\S$\ref{subsec:tidalstream} and relevant photometric features are labeled.}
	\end{figure} 

In addition to the large stellar loop discussed above, our deep wide-field images also reveal a plethora of faint diffuse light features that are identified in Figure \ref{fig:photofeatures} and are potentially associated with the ongoing destruction of an accreted dwarf galaxy. Recent simulations of 1:10 satellite-to-host mass ratio minor mergers by \citet{purcell2010} have shown that such accretion events can dynamically eject disk stars into a diffuse, azimuthally non-uniform stellar halo component with heights of $\gtrsim30$ kpc above the disk plane. As such, additional features of interest that are potentially related to the accretion event that resulted in the stellar loop are selected based on their color similarity to feature $a$ (and therefore also the inner stellar halo and outer disk of M63) as well as their location and spatially asymmetric nature.  

Feature $b$ is an enormous protrusion appearing to begin where feature $a$ enters the galaxy's outer isophotes and wrapping around the southeastern edge, ending in a surface brightness discontinuity to the south, southwest of the galaxy where feature $e$ begins. Feature $b$ is slightly brighter than $a$, but displays a very similar color of $B-R$ = 1.5$\pm$0.1 and corresponds to the redder edge of M63's light distribution, which was alluded to by \citet{vanderKruit1979} and can be seen in Figure \ref{fig:Colormap}. 

Feature $c$ is a curious asymmetry protruding slightly to the north at a smaller galactocentric radius than features $a$ or $b$. It is in close proximity to another location of slightly redder colors on the $B-R$ color index map in Figure \ref{fig:Colormap} and is significantly brighter than feature $a$ (probably due to the contribution of M63's faint outer light at the closer location of $c$ to the galaxy). Its measured color is consistent with M63's outer disk isophotes and the stream within the measurement errors.

Feature $e$ is an enormous, dim plume that has similar $\mu_{B}$ and $\mu_{R}$ as the dim break in feature $a$ and is located at a similar galactocentric radius. Its $B-R$ color is also consistent with the tidal debris and the outer disk isophotes. However, there is no HI gas associated with this feature down to the detection limit of Figure \ref{fig:HImap}. As such, this feature could be a part of the stellar tidal stream, but we refrain from making any definitive conclusions about it given that the photometric uncertainties are relatively large. 

To the west of the galaxy disk, it is possible to see a ``wing''-like feature labeled $f$. Like the features above, $f$ has a color similar to the outer disk and stellar halo as measured at the $\mu_{R}$ = 23.5 mag arcsec$^{-2}$ isophote. Although this giant plume could be a component of the tidal debris, it may instead be associated with an extension of a spiral arm into the outer disk (note that the HI spiral structure closely aligned with this feature extends up to a galactocentric radius of $\sim$40 kpc; \citealp{battaglia2006}).

Feature $d$ is an extremely dim extension of light towards the south of the galaxy and is visible only after very extreme non-linear stretches of our images. Since it is below our limit for accurate measurements, only a very uncertain surface brightness can be reported, probably around 1 mag arcsec$^{-2}$ dimmer than feature $a$. Since it appears in both the MDO $R$ band \textit{and} G-NMS images as a loop-like structure, it is unlikely to be a sky subtraction residual. The models of \citet{johnston2008} show that the remnants of accreted dwarfs dim as a function of time since the initial disruption as the satellite's stellar distribution becomes more diffuse. This implies that feature $d$ may be a much more ancient merger fossil. A similar dim loop-like structure can also be seen on the northern side of the galaxy extending towards the $V$ = 9.8 mag star SAO 44528. This bright star hinders our ability to study this possible northern component of the dim loop. 

Our data also detect the optical counterparts of long, clumpy filaments (features $g$ and $h$) clearly visible in GALEX ultra-violet (UV: $1350 < \lambda$ (\AA) $< 2800$) images of M63 \citep{gildepaz2007a}, reproduced here from archival data in Figure \ref{fig:GALEXmap}. As with the HI map, there are no hints of the stream (feature $a$) in these UV data, even though it displays a surface brightness similar to that of $g$ and $h$ in our optical images. Given that they are seen in both the HI map and GALEX imaging, this may suggest that $g$ and $h$ are composed of a younger stellar population than the stream and are ``patchy'' components of the XUV disk classified by \citet{thilker2007}. This is upheld by the blue color indices of these features ($B-R \approx$ 0.9). In Figure \ref{fig:GALEXmap}, we show the $\mu_{FUV} = 27.25$ AB mag arcsec$^{-2}$ isophote from \citet{thilker2007}. Patchy UV emission outside of this isophote helps to define the Type 1 XUV disks, a classification to which the M63 system belongs. As mentioned, these filaments can also be associated with parts of the outer spiral structure visible in the HI map in Figure \ref{fig:HImap}, unlike the loop feature and other suspected tidal debris. Embedded near feature $g$ is the large, bright clump of XUV emission, which looks to be a part of the XUV disk according to the GALEX data. However, this clump carries a UGCA (342) and PGC (46093) designation and is classified as a Magellanic Irregular galaxy by \citet{deVaucouleurs1991} and tidal dwarf galaxy by \citet{thilker2007}. \citet{bremnes1999} deduced from deep photometry that it was part of the M63 system rather than a separate galaxy. Given its blue $B-R$ color of 0.87$\pm$0.04 and its apparent association with XUV disk features in the GALEX image, we tend to agree with this claim. Thus, we do not consider UGCA 342 to be a viable candidate for the progenitor dwarf galaxy of the tidal stream, although kinematical data from optical spectroscopy would be useful to at least confirm if it is part of M63.	

\section{CONCLUSIONS AND DISCUSSION}\label{sec:Discussion}

\subsection{Origin of Faint Light Features}\label{subsec:Origin}
We present surface photometry of a stellar tidal stream and a deep panoramic view of a complex of substructure in the outer regions of the nearby spiral galaxy M63. Our data, collected independently from three different small telescopes, reveals an enormous arc-like structure around the galaxy's disk extending $\sim$29 kpc projected from its center, tilted with respect to its strong HI gaseous warp and the stellar disk. This strong indication of a recent merger event provides yet another example (in addition to NGC 5907 and NGC 4013) of an apparently isolated galaxy with a significantly warped gaseous disk that also shows clear evidence of the ongoing tidal disruption of a dwarf companion. Moreover, the appearance of the enormous great-circle arc feature that has no UV or HI counterpart is consistent with tidal debris resulting from ongoing satellite accretion in the $\Lambda$CDM based models of \citet{bullock2005}, \citet{johnston2008}, and \citet{cooper2010}. 

The sky-projected geometry of the stream provided by our images gives some insight to its possible origin. The large loop-like feature around M63 appears to belong to the great-circle type stellar tidal streams (like the tidal stream of the Sagittarius dwarf and that around NGC 5907). From the analysis of $\sim$1500 accretion events in 11 stellar halos that are hierarchically constructed in a $\Lambda$CDM universe, \citet{johnston2008} find that unbound great-circle morphologies (i.e., those with no remaining bound core of particles) arise from accretion events typically beginning 6-10 Gyr ago on near circular orbits. These simulations also find that this is the predominant morphological type associated with still-bound satellites which arise from more recent events (i.e., within the last 6 Gyr) on mildly eccentric orbits. Given the measured surface brightness of $\mu_{R} \approx$ 26.1 mag arcsec$^{-2}$ for the M63 stream, the accretion event considered here would fall near the brightest of those in the \citet{johnston2008} models, indicating that the progenitor satellite may have been accreted in the last several Gyr and may still be bound. A discussion of the prospects for a remaining still-bound satellite core is given in the following subsection.

Our observations of M63 are consistent with counts of substructure in cosmological simulations for the surface brightness limit of our pilot survey \citep{bullock2005, johnston2008, cooper2010, martinezDelgado2010}. \citet{johnston2008} conclude that finding a single satellite clearly in the process of disruption with debris spread around the host galaxy should not be surprising in a survey with our sensitivity, but finding many satellites in that state around a single galaxy would be unusual. Yet, given the existence of extremely faint features (such as feature $d$) that are below our reasonable measurement limit (i.e., the derived uncertainties are $\sim0.5-1.0$ mag arcsec$^{-2}$) and at least 1 mag arcsec$^{-2}$ dimmer than feature $a$, for example, it is possible that we have observed two such accretion events that occurred at different epochs.  

From the presented data, only feature $a$ can convincingly be determined to be a part of a coherent stellar stream. However, our data also feature a number of other faint light substructures around M63, as listed in the previous section. While we have no conclusive evidence that any of these other features are additional components of the same coherent stellar stream, there is reason to believe that they may be ``side-effects'' resulting from the same accretion event's influence on the M63 parent system. These possibilities are discussed below in $\S$\ref{subsec:ParentEffect}.

\subsection{Location and Fate of the Progenitor Dwarf Galaxy}\label{subsec:RemainingDwarf}
Our images do not provide any obvious insight on the current position or final fate of the progenitor dwarf galaxy of this stream. The field of the G-NMS image includes a satellite, UGC 8313 (see Figure \ref{fig:photofeatures}), a galaxy of type SB(s)c \citep{deVaucouleurs1991} and another member of the M51 galaxy group \citep{fouque1992} at a projected distance of $\sim$50 kpc to the northwest. Although our deep image reveals some hint of a possible faint stellar warp on the northern edge of UGC 8313, the galaxy's position and lack of any tidal debris in its vicinity make it very unlikely that this satellite is related to the main stream to the northeast. As mentioned in $\S$\ref{subsec:otherfeatures}, the nearby ``condensation" UGCA 342 is also likely not to be a viable candidate for the missing progenitor satellite given its apparent alignment with the XUV disk discovered by the GALEX satellite \citep{thilker2007} and its extremely different $B-R$ color as compared to the stellar stream (see Table \ref{table:PhotoData}). 

The surviving satellite is likely indiscernible inside the stream due to a currently small size and especially low surface brightness, which decreases monotonically once stellar material begins to be stripped from the progenitor dwarf in the parent's tidal field (see $\S$5 of \citealp{martinezDelgado2008} for a discussion on this issue). As in our previously studied tidal streams \citep{martinezDelgado2008, martinezDelgado2009}, a surviving satellite could alternatively be hidden behind or superposed on the spiral disk. Because of M63's lower inclination as compared to these examples, the probability of this possibility is increased. Although the isophotes of Figure \ref{fig:SBmap} show no obvious corresponding light over-density, there is a small area of especially red $B-R$ color to the north of the masked portion of M63's disk near where feature $a$ is lost in the outer disk light, east of feature $c$ (see Figure \ref{fig:Colormap}). Although the data we have at hand cannot confirm nor refute it, this redder patch could potentially be the low surface brightness remains of a surviving dwarf galaxy core that is nearly washed out by the faint disk light due to its projected proximity and current low surface brightness. By a similar reasoning, the remains could be hidden in feature $b$, which also has a red $B-R$ color and is located close to the outer disk in projection.  

Alternatively, it is possible that the progenitor satellite is totally disrupted at the present time. Such a scenario would imply an old, multiply wrapped stellar stream since total disruption is unlikely on a single passage of a satellite on a near circular orbit (e.g., see Figure 3 of \citealp{penarrubia2010b}). However, as discussed in $\S$\ref{subsec:nLoops}, our data do not provide any conclusive evidence for a long, multiply wrapped stellar stream thanks in large part to the difficulty of observing faint features in close vicinity to the moderately inclined outer disk of M63.

\subsection{Mass and Age of the Stellar Stream}\label{subsec:MassAge}
The observed morphology of feature $a$ suggests a great-circle, rosette-like stellar tidal stream (where the satellite and its corresponding debris are in a near circular orbit). Thus, we can give illustrative estimates of the mass of the disrupted satellite galaxy as well as the time since disruption using the analytical formalism of \citet{johnston2001}. This procedure was adopted by \citet{johnston2001} for the stream around NGC 5907 and by \citet{wehner2005} for the stream around NGC 3310. Equations 12 and 13 of \citealp{johnston2001} for mass and age,\footnote{Note that age in this context is the time since disruption, not the age of the stellar population. Our observations of $B-R$ alone cannot constrain the stellar population of the tidal stream.} respectively, are valid for disruption times of $t < 3T_{\Psi} = 6\pi R / v_{circ}$, where $T_{\Psi}$ is the orbital period of the satellite and $R$ is the orbit's typical radius. This time is $\sim$3.0 Gyr for the system in consideration here (using a rough estimate for the stream radius $R = 29$ kpc from feature $a$'s projected maximum distance and $v_{circ} \approx 180$ km s$^{-1}$ at $R$ from \citealp{battaglia2006}). To estimate the mass, we use the stream width $w$ = 3.3 kpc (see \S\ref{subsec:tidalstream}), which as noted does not appear to significantly vary along its length. This may be due to the contamination of faint light from the galaxy, which also inhibits our ability to determine the pericentric distance of the orbit $R_{P}$. However, the lack of variability in $w$ may suggest that the time since the satellite's disruption is small (i.e., $t < 1.5T_{\Psi}$ from \citealp{johnston2001}, $\S$2.2). Since the projected view of the galaxy and the stream does not yield an obvious measure of $R_{P}$, we simply assume an orbit with small eccentricity such that $R_{P} \approx R$ (this is not a bad approximation for streams having the great-circle morphology). Using this information and Equation 12 of \citet{johnston2001}, we estimate the mass of the progenitor satellite to be $\sim$3.5$\times$10$^{8}$ M$_{\odot}$. This is of the same order as the significant progenitor satellites which assemble stellar halos from 1 $< z <$ 7 in the simulations of \citet{cooper2010} and of the same order as the total mass of several Local Group dwarf spheroidals, such as NGC 147, NGC 185, and the Fornax Dwarf \citep{mateo1998}. 

From Equation 13 of \citet{johnston2001}, we can see that the estimate of the time since disruption depends linearly on $\Psi$, the angular length of the stream. As will be discussed in the following section, the true value of $\Psi$ is not conclusive from our data. Therefore, we parameterize the stream's angular length as $\Psi = 2\pi\eta$, where $\eta$ is thus the total number of wraps the stream makes around M63. Using the radius of a circular orbit with the same energy as the true orbit $R_{circ} = (R_{A} + R_{P}) / 2 \approx R$ (where $R_{A}$ is the apocenter distance), we estimate that $t \approx 1.8\eta$ Gyr. While large uncertainty exists in these illustrative calculations, a reasonable value of $\eta$ shows that the results obtained using the analytical approach of \citet{johnston2001} and the results of the more recent modeling of \citet{johnston2008} are consistent with an accretion event having occurred in the last few Gyr. 

As an alternative to the dynamical approach above, we can also estimate the mass of the progenitor satellite by measuring the stream's $R$ band luminosity density $\Sigma_{R}$ and estimating a mass-to-light ratio $M/L_{R}$. This method was used to estimate the mass of the progenitor dwarf of the NGC 5907 stellar stream \citep{martinezDelgado2008} and to measure surface mass density profiles of spiral galaxies by \citet{bakos2008}. Since we can clearly make a distinction of only feature $a$ from the outer faint disk light of M63, we estimate $\Sigma_{R}$ using its average value of $\mu_{R} = 26.1$ mag arcsec$^{-2}$. With the absolute $R$ magnitude of the sun\footnote{See: \textit{http://mips.as.arizona.edu/$\sim$cnaw/sun.html}} $M_{R \odot} = 4.46$, we can arrive at the well-known expression:
\begin{equation}
\Sigma_{R} = 10^{0.4 (26.03 - \mu_{R})} \left(\frac{L_{R \odot}}{pc^{2}}\right)\;,
\end{equation}   
yielding $0.94^{+0.09}_{-0.08}$ $L_{R \odot}$ pc$^{-2}$. Using the result of the ellipse fitting of feature $a$ in $\S$\ref{subsec:tidalstream}, we estimate the projected area $A$ subtended by the stream as: 
\begin{equation}
A = \eta \pi q \left[\left(a + \frac{w}{2}\right)^2 - \left(a - \frac{w}{2}\right)^2\right]\;.
\end{equation}
This results in $A = 223 \eta$ kpc$^2$. The total $R$ band luminosity is then $\Sigma_{R} A = (2.1\pm0.2) \eta \times 10^8$ $L_{R \odot}$. \citet{belldeJong2001} have used stellar population synthesis models to predict the dependence of $M/L$ on broadband colors. Using Table 1 of that work (i.e., $a_{R} = 0.820$ with a 0.15 dex reduction for a Kroupa IMF and $b_{R} = 0.851$) and our estimation of the stream $B-R$ color, we estimate that $M/L_{R} = 2.0^{+1.0}_{-0.7}$. This yields a total mass for the stream of $(4\pm2)\eta \times 10^8$ $M_{\odot}$.
 
\subsection{Length of the Tidal Stream}\label{subsec:nLoops}
Our deep observations of M63 only clearly reveal a single coherent arc (feature $a$) having an angular length of $\Psi \approx \pi$, which yields a lower limit of $\eta \gtrsim 0.5$. This implies that $t \gtrsim 0.9$ Gyr. However, one can immediately see that $\eta \approx 1$ allows the two mass determinations from the previous section to agree quite well at $\sim 4 \times 10^8$ $M_{\odot}$.

In $\S$\ref{subsec:tidalstream}, we pointed out a break in the surface brightness distribution of the otherwise coherent arc. Recent work by \citet{yoon2011} has shown that such breaks in cold stellar streams can be caused by a direct impact with another massive body. Given the scale of this accretion event and the angular size of the gap ($\sim$10$^{\circ}$ projected), the impacting body would probably need to be $\gtrsim 10^{8}$ $M_{\odot}$ (see Figure 5 of \citealp{yoon2011} which compares the streams of Palomar 5 and the Sagittarius dwarf). Since subhalos of $\sim 10^8$ $M_{\odot}$ are thought to contain stars (e.g., \citealp{greif2008}), the impacting body could be another accreted dwarf galaxy (or the remnants thereof; here, perhaps the dwarf associated with feature $d$). Alternatively, gaps in stellar streams can occur naturally for satellites on fairly elliptical orbits. This is because mass-loss occurs variably about the orbit with most occurring abruptly near perigalacticon, as illustrated in Figure 3 of \citet{penarrubia2010b}. We cannot rule out this possibility for explaining the stream's gap because as previously mentioned, the projection of the stream on the sky plane is currently unknown. Thus, we cannot determine the eccentricity of the orbit using our images without additional kinematic data.  

Besides the above interesting possibilities, the dim break in the coherent stream could provide a clue for estimating $\Psi$. For example, the break could simply be a gap between the ends of the leading and trailing arms of the tidal stream as the first orbital wrap since disruption is being completed. This interpretation seems plausible given that if the interaction began recently enough to have formed only a single loop at the present time, we might expect to detect the remaining progenitor satellite \textit{in} the stream since it would be midway between the leading and trailing streamers \citep{johnston2001}. Given the position of the gap, the remaining core of the satellite might then be hidden behind or superposed on the disk (thus evading clear detection; see $\S$\ref{subsec:RemainingDwarf}). Such a scenario could also explain the systematic red color gradient observed on the east side of the outskirts of M63 (see Figure \ref{fig:Colormap}): given a single wrap, the redder $B-R$ colors on the east side of the galaxy could result from the stream being above the disk plane in the line of sight with the opposite explaining the lack of red $B-R$ color indices on the west side. If this is the case, $\eta \approx 1$ and $t \approx 1.8$ Gyr. 

\citet{johnston2001} addressed the difficulty of observing faint features in the outskirts of spiral galaxies that are not seen edge-on. Since M63 has $i_{M63}$ $\approx$ 53.3$^{\circ}$, it is far from edge-on and faint features at a height of up to $\sim$20 kpc above the galaxy plane can become difficult to discern. Thus, there is no clear evidence for $\eta > 1$ because our data are not conclusive about whether other faint features (such as $b$, $c$, $e$, or $f$) are a part of the stellar stream or are instead native to M63 itself. 

\subsection{Effect on the Parent Galaxy System}\label{subsec:ParentEffect}
It is interesting to consider in what ways a minor merger such as the one we have observed might affect the parent galaxy system since such events should be common. Our view of the ongoing disruption of a satellite galaxy around M63 could hint at the inside-out formation of the stellar halo. As can be seen through our photometry (Table \ref{table:PhotoData}), the stellar halo's color is fairly red and quite similar to the color of the stellar stream, indicating that it consists of a mix of stars. As shown by recent simulations by \citet{purcell2010}, accreted stars from the disrupted dwarf can occupy the same outer regions as the thick envelope of stars ejected from the outer disk as a result of the interaction. While our data hint at a systematic color gradient towards redder $B-R$ colors near the stream (see Figure \ref{fig:Colormap}), the simulations imply that it should not be surprising to find a similar color between the stream and outer regions of M63. 

Such a merger could also explain some of the other low surface brightness features we observe around M63, such as features $b$, $c$, $e$, and $f$. The simulations of \citet{purcell2009, purcell2010}, which model 1:10 satellite-to-host mass ratio minor mergers with a variety of orbital inclinations, show that large groups of stars can be ejected from the stellar disk to large heights ($\gtrsim 30$ kpc) in an azimuthally asymmetric manner. As seen in the simulations, such stellar disturbances in the outer disk could be an explanation for the large ``wing''-like features mentioned. We note that our surface photometry shows that all of the listed photometric features' $B-R$ colors (except for $b$) are fully consistent with being \textit{the same} color as the outer disk as measured at the $\mu_{R} = 23.5$ mag arcsec$^{-2}$ isophote, which supports this claim. Especially convincing are features $c$ and $f$ given their small photometric uncertainties. After several Gyr, such structures likely settle into a thick disk component.

As has already been mentioned, M63 is the most recent of several galaxies (e.g., NGC 5097 and NGC 4013) discovered to have an extremely warped HI disk in the presence of a recent merger event. These examples counter the argument of \citet{sancisi1976} who dismiss the tidal origin of such warps due to the assumption of galaxy isolation. The interaction may also affect the HI disk in other ways. As pointed out in $\S$\ref{subsec:tidalstream}, there are two enormous voids (each $\gtrsim15$ kpc in size) in the HI distribution, spaced roughly $180^{\circ}$ apart and located roughly where the tidal stream intersects the disk. Simulations by \citet{bekkiChiba2006} and \citet{kannan2011} have shown that an impact of a dark subhalo of $10^{8}$ $M_{\odot}$ on the HI disk can create holes on kpc scales. These simulations require the halos to contain some modest gas fraction to supply a dissipational force to create a void in the gas, since gravity alone can cause HI \textit{over}-densities due to gravitational focusing \citep{kannan2011}. Given that the stellar stream appears to have no gas associated with it, and the $B-R$ color and estimated mass are consistent with gas-poor dwarf spheroidals and S0 galaxies, it would seem unlikely that there would be enough gas in the progenitor satellite galaxy to produce HI voids of such large scale upon impact. Additionally, the cited simulations produce rings of high density gas around the void, which induces star formation. We do not observe enhanced star formation around the voids either in our deep optical images or in the GALEX data. Finally, in the time it would take the satellite to complete half of an orbit, the rotation of the HI disk would have likely rotated the voids out of alignment with the stream such that the collision points on the disk would not be located in such a $180^{\circ}$ symmetric fashion. 

As an alternative explanation, companions as little as 1\% of the mass of the parent galaxy have been theorized to drive tidally induced spiral structure \citep{byrdhoward1992,mihoshernquist1994}. As such, the ongoing minor merger studied in this work could possibly be an explanation for the underlying $K$ band spiral structure in M63 observed by \citet{thornley1996}. Similarly, the extension of these tidally driven spiral arms into the outer disk could induce instabilities in the gaseous disk at high radii giving rise to Type 1 XUV disk features outside of where traditional star formation occurs. As simulated by \citet{bush2008}, this scenario is plausible because inner spiral structure propagates into the gaseous HI outer disk. Indeed, such spiral over-densities of gas can be seen in the HI map (Figure \ref{fig:HImap}; \citealp{battaglia2006}) corresponding to the location of XUV emission detected both in our deep optical images (e.g., features $g$ and $h$) and in GALEX imaging (Figure \ref{fig:GALEXmap}). The propagation of the spiral density waves at high radii in the HI disk could also be responsible for clearing out the large-scale voids.

\citet{thilker2007} looked for a link between interactions and XUV disks. While their statistical tests comparing the mean perturbation parameter $f$ \citep{varela2004} for the entire sample to that for galaxies hosting a Type 1 XUV disk morphology showed no \textit{significant} link, those authors do admit some limitations. In particular, the $f$ parameter does not take into account undetected low surface brightness objects (or in this case, debris or faint stellar streams). \citet{thilker2007} also find that 75\% of Type 1 XUV disk objects show morphological evidence (including HI warps) of an interaction, merger, or external perturbation and suggest that interactions may still be a viable way to drive star formation in the outer disk. M63 was classified as relatively isolated in that work, and in light of the ongoing interaction presented here, this may require re-evaluation. Indeed, other galaxies around which stellar tidal streams have been discovered also display Type 1 XUV disks. In particular, M83 shows XUV disk emission also indicating star formation at extremely high radii \citep{gildepaz2007b}, has a value of $f$ indicating isolation \citep{thilker2007}, and yet displays at least one great-circle like tidal stream indicative of an ongoing interaction with a satellite \citep{malin1997}. Exploration of a larger sample of galaxies showing previously undetected evidence of a disrupted dwarf satellite that is also common to the sample of \citet{thilker2007} is necessary to further explore the possible link between minor mergers and XUV disks. 

With the increasing number of discoveries of stellar halo substructure and streams around external galaxies (which will likely grow quickly thanks to a new systematic survey for such features; \citealp{martinezDelgado2010}), we are learning that satellite accretion is an essential part of galaxy evolution. The near future will likely bring new hypotheses about the impact of such events on the evolution of spiral galaxies, including disk warps or extended star formation, which can be based on reliable statistics rather than speculation based on a few individual systems. The results of our pilot survey of stellar tidal streams in nearby galaxies \citep{martinezDelgado2010}, including the deep images of M63 and its tidal stream analyzed in this work, have shown how the stellar halos of spiral galaxies in the Local Universe still contain a significant number of relics from their hierarchical, inside-out formation. This presents us with a unique opportunity to be witnesses of the latest stages of galaxy evolution and the ongoing assembly of spiral galaxy halos.\\

This work is partially supported by the Texas Norman Hackerman Advanced Research Program under grant 003658-0295-2007. S.R.M. appreciates support from NSF grant AST-0807945. We would like to thank McDonald Observatory and its staff for supporting the photometric observations, T. Taylor for useful bits of code which aided in the analysis, A. Gil de Paz, M. A. G\'{o}mez-Flechoso, N. Martin, E. Bell, H. W. Rix, S. Odewahn, M. Cornell, and K. Gebhardt for useful scientific discussions, K. Jordi for supplying the machine readable catalog of stars with matched $BVRI$ and $ugriz$ photometry, and G. Battaglia for providing the electronic format of the HI map of M63. Additionally, we would like to acknowledge the anonymous referee for useful comments and suggestions which have improved this paper. 

Funding for the SDSS has been provided by the Alfred P. Sloan Foundation, the Participating Institutions, the National Science Foundation, the U.S. Department of Energy, the National Aeronautics and Space Administration, the Japanese Monbukagakusho, the Max Planck Society, and the Higher Education Funding Council for England. The SDSS Web Site is \textit{http://www.sdss.org/}. GALEX \textit{(Galaxy Evolution Explorer)} is a NASA Small Explorer, launched in 2003 April. We acknowledge NASA's support for construction, operation, and science analysis for the GALEX mission, developed in cooperation with the Centre National d`Etudes Spatiales of France and the Korean Ministry of Science and Technology. This research has made use of the NASA/IPAC Extragalactic Database (NED) and the NASA/IPAC Infrared Science Archive, which is operated by the Jet Propulsion Laboratory, California Institute of Technology, under contract with the National Aeronautics and Space Administration. We acknowledge the use of the HYPERLEDA database (\textit{http://leda.univ-lyon1.fr}).  
\\




\begin{thebibliography}{}

\bibitem[Abazajian et al.(2009)]{abazajian2009} Abazajian, K.N. et al. 2009, \apjs, 182, 543

\bibitem[Adelman-McCarthy et al.(2006)]{adelmanmccarthy2006} Adelman-McCarthy, J.K. et al. 2006, \apjs, 162, 38

\bibitem[Bakos et al.(2008)]{bakos2008} Bakos, J., Trujillo, I., \& Pohlen, M. 2008, \apjl, 683, L103

\bibitem[Barway et al.(2005)]{barway2005} Barway, S., Mayya, Y.D., Kembhavi, A.K., \& Pandey, S.K. 2005, \aj, 129, 630

\bibitem[Battaglia et al.(2006)]{battaglia2006} Battaglia, G., Fraternali, F., Oosterloo, T., \& Sancisi, R. 2006, \aap, 447, 49

\bibitem[Bekki \& Chiba(2006)]{bekkiChiba2006} Bekki, K. \& Chiba, M. 2006, \apjl, 637, L97

\bibitem[Bell \& de Jong(2001)]{belldeJong2001} Bell, E.F. \& de Jong, R.S. \apj, 550, 212

\bibitem[Belokurov et al.(2006)]{belokurov2006} Belokurov, V. et al. 2006, \apjl, 642, L137

\bibitem[Belokurov et al.(2007)]{belokurov2007} Belokurov, V. et al. 2007, \apj, 658, 337

\bibitem[Bremnes et al.(1999)]{bremnes1999} Bremnes, T., Binggeli, B., \& Prugniel, P. 1999, \aaps, 137, 337

\bibitem[Bullock \& Johnston(2005)]{bullock2005} Bullock, J.S. \& Johnston, K.V. 2005, \apj, 635, 931

\bibitem[Bush et al.(2008)]{bush2008} Bush, S. J., Cox, T.J., Hernquist, L., Thilker, D., \& Younger, J.D. 2008, \apjl, 683, L13

\bibitem[Byrd \& Howard(1992)]{byrdhoward1992} Byrd, G. G. \& Howard, S. 1992, \apj, 103, 1089

\bibitem[Chonis \& Gaskell(2008)]{chonisGaskell2008} Chonis, T.S. \& Gaskell, C.M. 2008, \aj, 135, 264
 
\bibitem[Claver(1992)]{claver1992} Claver, C.F. 1992, BAAS, 24, 1282

\bibitem[Cooper et al.(2010)]{cooper2010} Cooper, A.P., et al. 2010, \mnras, 406, 744

\bibitem[de Vaucouleurs et al.(1991)]{deVaucouleurs1991} de Vaucouleurs, G., de Vaucouleurs, A., Corwin, Jr., H.G., Buta, R.J., Paturel, G., \& Fouque, P. 1991, {\it Third Reference Catalog of Bright Galaxies}, Springer [New York] (RC3)

\bibitem[Elmegreen \& Elmegreen(1987)]{elmegreen1987} Elmegreen, D. M. \& Elmegreen, B. G. 1987, \apj, 314, 3

\bibitem[Fouque et al.(1992)]{fouque1992} Fouque, P., Gourgoulhon, E., Chamaraux, P., \& Paturel, G. 1992, \aaps, 93, 211

\bibitem[Gil de Paz et al.(2007a)]{gildepaz2007a} Gil de Paz, A. et al. 2007a, \apjs, 173, 185  

\bibitem[Gil de Paz et al.(2007b)]{gildepaz2007b} Gil de Paz, A. et al. 2007b, \apj, 661, 115

\bibitem[Greif et al.(2008)]{greif2008} Greif, T.H., Johnson, J.L., Klessen, R.S., \& Bromm, V. 2008, \mnras, 387, 1021

\bibitem[Grillmair(2006)]{grillmair2006} Grillmair, C.J. 2006, \apjl, 651, L29

\bibitem[Hernquist \& Quinn(1993)]{hernquistquinn1993} Hernquist, L. \& Quinn, P. 1993, \textit{ASPC}, 49, 187

\bibitem[Ibata et al.(2001a)]{ibata2001} Ibata, R., Irwin, M., Lewis, G., Ferguson, A.M.N., \& Tanvir, N. 2001a, \nat, 412, 49-52 

\bibitem[Ibata et al.(2001b)]{ibata2001b} Ibata, R., Lewis, G.F., Irwin, M., Edward, T., \& Quinn, T. 2001b, \apj, 551, 294

\bibitem[Ibata et al.(2007)]{ibata2007} Ibata, R., Martin, N.F., Irwin, M., Chapman, S., Ferguson, A.M.N., Lewis, G.F. \& McConnachie, A.W. 2007, \apj, 671, 1591-1623

\bibitem[Jansen et al.(2000)]{jansen2000} Jansen, R.A., Franx, M., Fabricant, D., \& Caldwell, N. 2000, \apjs, 126, 271

\bibitem[Johnston et al.(2001)]{johnston2001} Johnston, K.V., Sackett, P.D., \& Bullock, J.S. 2001, \apj, 557, 137

\bibitem[Johnston et al.(2008)]{johnston2008} Johnston, K.V., Bullock, J.S., Sharma, S., Font, A., Robertson, B.E., \& Leitner, S.N. 2008, \apj, 689, 936 

\bibitem[Jordi et al.(2006)]{jordi2006} Jordi, K., Grebel, E.K., \& Ammon, K. 2006, \aap, 460, 339

\bibitem[Kannan et al.(2011)]{kannan2011} Kannan, R., Macci\'{o}, A.V., Pasquali, A., Moster, B.P., \& Walter, F. 2011, arXiv:1107.2401

\bibitem[Law \& Majewski(2010)]{law2010} Law, D.R. \& Majewski, S.R. 2010, \apj, 714, 229

\bibitem[Majewski et al.(2003)]{majewski2003} Majewski, S. R., Skrutskie, M., Weinberg, M., \& Ostheimer, J. 2003, \apj, 599, 1082

\bibitem[Malin \& Hadley(1997)]{malin1997} Malin, D. \& Hadley, B. 1997, Publ. Astron. Soc. Australia, 14, 52

\bibitem[Martin \& Kennicutt(2001)]{martin2001} Martin, C.L. \& Kennicutt, Jr., R.C. 2001, \apj, 555, 301

\bibitem[Mart\'{i}nez-Delgado et al.(2004)]{martinezDelgado2004} Mart\'{i}nez-Delgado, D., G\'{o}mez-Flechoso, M.A., Aparicio, A., \& Carrera, R. 2004, \apj, 601, 242

\bibitem[Mart\'{i}nez-Delgado et al.(2008)]{martinezDelgado2008} Mart\'{i}nez-Delgado, D., Pe\~{n}arrubia, J., Gabany, R.J., Trujillo, I., Majewski, S.R., \& Pohlen, M. 2008, \apj, 689, 184

\bibitem[Mart\'{i}nez-Delgado et al.(2009)]{martinezDelgado2009} Mart\'{i}nez-Delgado, D., Pohlen, M., Gabany, R.J., Majewski, S.R., Pe\~{n}arrubia, J. \& Palma, C. 2009, \apj, 692, 955

\bibitem[Mart\'{i}nez-Delgado et al.(2010)]{martinezDelgado2010} Mart\'{i}nez-Delgado, D. et al. 2010, \aj, 140, 962

\bibitem[Mateo(1998)]{mateo1998} Mateo, M. 1998, \araa, 36, 435

\bibitem[McConnachie et al.(2009)]{mcconnachie2009} McConnachie, A.W. et al. 2009, \textit{Nature}, 461, 66

\bibitem[McConnachie et al.(2010)]{mcconnachie2010} McConnachie, A.W., Ferguson, A.M.N., Irwin, M.J., Dubinski, J., Widrow, L.M., Dotter, A., Ibata, R., \& Lewis, G.F. 2010, \apj, 723, 1038

\bibitem[Mihos \& Hernquist(1994)]{mihoshernquist1994} Mihos, J.C. \& Hernquist, L. 1994, \apjl, 425, L13

\bibitem[Moster et al.(2010)]{moster2010} Moster, B.P., Macci\'{o}, A.V., Somerville, R.S., Johansson, P.H., \& Naab, T. 2010, \mnras, 403, 1009

\bibitem[Mouhcine et al.(2010)]{mouhcine2010} Mouhcine, M., Ibata, R., \& Rejkuba, M. 2010, \apjl, 714, L12

\bibitem[Pe\~{n}arrubia et al.(2010a)]{penarrubia2010a} Pe\~{n}arrubia, J., Belokurov, V., Evans, N.W., Mart\'{i}nez-Delgado, D., Gilmore, G., Irwin, M., Niederste-Ostholt, M. \& Zucker, D.B. 2010a, \mnras, 408, L26

\bibitem[Pe\~{n}arrubia et al.(2010b)]{penarrubia2010b} Pe\~{n}arrubia, J., Benson, A.J., Walker, M.G., Gilmore, G., McConnachie, A.W., \& Mayer, L. 2010b, \mnras, 406, 1290

\bibitem[Pierce(1994)]{pierce1994} Pierce, M. 1994, \apj, 430, 53

\bibitem[Purcell et al.(2009)]{purcell2009} Purcell, C.W., Kazantzidis, S., \& Bullock, J.S. 2009, \apjl, 694, L98 

\bibitem[Purcell et al.(2010)]{purcell2010} Purcell, C.W., Bullock, J.S., \& Kazantzidis, S. 2010, \mnras, 404, 1711

\bibitem[Robaina et al.(2009)]{robaina2009} Robaina, A.R. et al. 2009, \apj, 704, 324

\bibitem[Robertson et al.(2006)]{robertson2006} Robertson, B., Bullock, J.S., Cox, T.J., Di Matteo, T., Hernquist, L., Springel, V., \& Yoshida, N. 2006, \apj, 645, 986

\bibitem[Sancisi(1976)]{sancisi1976} Sancisi, R. 1976, \aap, 53, 159

\bibitem[Schlegel et al.(1998)]{schlegel1998} Schlegel, D.J., Finkbeiner, D.P., \& Davis, M. 1998, \apj, 500, 523

\bibitem[Searle \& Zinn(1978)]{searlezinn1978} Searle, L. \& Zinn, R. 1978, \apj, 225, 357

\bibitem[Shang et al.(1998)]{shang1998} Shang, Z. et al. 1998, \apjl, 504, L23

\bibitem[Sollima et al.(2010)]{sollima2010} Sollima, A., Gil de Paz, A., Mart\'{i}nez-Delgado, D., Gabany, R.J., Gallego-Laborda, J.J., \& Hallas, T. 2010, \aap, 516, 83 

\bibitem[Springel et al.(2005)]{springel2005} Springel, V. et al. 2005, \nat, 435, 629

\bibitem[Stetson(2000)]{stetson2000} Stetson, P.B. 2000, \pasp, 112, 925

\bibitem[Thilker et al.(2007)]{thilker2007} Thilker, D.A. et al. 2007, \apjs, 173, 538

\bibitem[Thornley(1996)]{thornley1996} Thornley, M.D. 1996, \apjl, 469, L45

\bibitem[Tonry \& Schneider(1988)]{tonry1988} Tonry, J. \& Schneider, D.P. 1988, \aj, 96, 807

\bibitem[Toomre \& Toomre(1972)]{toomre1972} Toomre, A. \& Toomre, J. 1972, \apj, 178, 623

\bibitem[T\'oth \& Ostriker(1992)]{tothostriker1992} T\'oth, G. \& Ostriker, J.P. 1992, \apj, 389, 5

\bibitem[Trujillo et al.(2009)]{trujillo2009} Trujillo, I., Mart\'{i}nez-Valpuesta, I., Mart\'{i}nez-Delgado, D., Pe\~{n}arrubia, J., Gabany, R.J., \& Pohlen, M. 2009, \apj, 704, 618

\bibitem[van der Kruit(1979)]{vanderKruit1979} van der Kruit, P.C. 1979, \aaps, 38, 15

\bibitem[Varela et al.(2004)]{varela2004} Varela, J., Moles, M., M\'{a}rquez, I., Galletta, G., Masegosa, J. \& Bettoni, D. 2004, \aap, 420, 873

\bibitem[Vel\'{a}zquez \& White(1999)]{velazquezwhite1999} Vel\'{a}zquez, H. \& White, S.D.M. 1999, \mnras, 304, 254

\bibitem[Wehner \& Gallagher(2005)]{wehner2005} Wehner, E.H., \& Gallagher, III, J.S. 2005, \apjl, 618, L21

\bibitem[Wehner et al.(2006)]{wehner2006} Wehner, E.H., Gallagher, J.S., Papaderos, P., Fritze-von Alvensleben, U., \& Westfall, K.B. 2006, \mnras, 371, 1047

\bibitem[White \& Frenk(1991)]{white1991} White, S. D. M. \& Frenk, C. S. 1991, \apj, 379, 52

\bibitem[Wild(1997)]{wild1997} Wild, W.J. 1997, \pasp, 109, 1269

\bibitem[Yanny et al.(2003)]{yanny2003} Yanny, B. et al. 2003, \apj, 588, 824

\bibitem[Yoon et al.(2011)]{yoon2011} Yoon, J.H., Johnston, K.V., \& Hogg, D.W. 2011, \apj, 731, 58

\bibitem[York et al.(2000)]{york2000} York, D.G. et al. 2000, \aj, 120, 1579

\bibitem[Zheng et al.(1999)]{zheng1999} Zheng, Z. et al. 1999, \aj, 117, 2757

\end{thebibliography}
\end{document}